\documentclass[12pt,a4paper]{article}
\usepackage[utf8]{inputenc}
\usepackage{color}
\usepackage{float}
\usepackage{mathtools}
\usepackage{amsmath}
\usepackage{amssymb}
\usepackage{cancel}
\usepackage{graphicx}
\PassOptionsToPackage{normalem}{ulem}
\usepackage{ulem}
\usepackage{cite}

\makeatletter

\pdfpageheight\paperheight
\pdfpagewidth\paperwidth

\providecommand{\tabularnewline}{\\}

\usepackage{url}\usepackage{esdiff}\usepackage{dsfont}\usepackage{appendix}\usepackage{caption}\usepackage{subcaption}\usepackage{footnote}\parskip = 12pt
\pdfoutput=1

\def\AA{\mathrel{%
   \rlap{\hspace{0.45ex}\raise 0.7ex \hbox{$^\circ$}}{\rm A}}}

\def\~{{$\tilde{\phantom{a}}$}}

\def\lsim{\mathrel {\vcenter {\baselineskip 0pt \kern 0pt
    \hbox{$<$} \kern 0pt \hbox{$\sim$} }}}
\def\gsim{\mathrel {\vcenter {\baselineskip 0pt \kern 0pt
    \hbox{$>$} \kern 0pt \hbox{$\sim$} }}}
\def\gtlt{\mathrel {\vcenter {\baselineskip 0pt \kern 0pt
    \hbox{$>$} \kern 0pt \hbox{$<$} }}}

\def\hline{\noalign{\hrule \vskip2pt}}

\def\|{\ifmmode\Vert\else \char`\|\fi}
\ifx\oldzeta\undefined                          
  \let\oldzeta=\zeta                            
  \def\zzeta{{\raise 2pt\hbox{$\oldzeta$}}}     
  \let\zeta=\zzeta                              
\fi

\ifx\oldchi\undefined                           
  \let\oldchi=\chi                              
  \def\cchi{{\raise 2pt\hbox{$\oldchi$}}}       
  \let\chi=\cchi                                
\fi

\def\frac#1#2{{#1 \over #2}}

\def\half{\ifinner {\scriptstyle {1 \over 2}}
   \else {1 \over 2} \fi}

\def\simge{\mathrel{%
   \rlap{\raise 0.511ex \hbox{$>$}}{\lower 0.511ex \hbox{$\sim$}}}}
\def\simle{\mathrel{
   \rlap{\raise 0.511ex \hbox{$<$}}{\lower 0.511ex \hbox{$\sim$}}}}

\def\buildchar#1#2#3{{\null\!                   
   \mathop#1\limits^{#2}_{#3}                   
   \!\null}}                                    
\def\overcirc#1{\buildchar{#1}{\circ}{}}

\def\slashchar#1{\setbox0=\hbox{$#1$}           
   \dimen0=\wd0                                 
   \setbox1=\hbox{/} \dimen1=\wd1               
   \ifdim\dimen0>\dimen1                        
      \rlap{\hbox to \dimen0{\hfil/\hfil}}      
      #1                                        
   \else                                        
      \rlap{\hbox to \dimen1{\hfil$#1$\hfil}}   
      /                                         
   \fi}                                         %

\def\subrightarrow#1{
  \setbox0=\hbox{
    $\displaystyle\mathop{}
    \limits_{#1}$}
  \dimen0=\wd0
  \advance \dimen0 by .5em
  \mathrel{
    \mathop{\hbox to \dimen0{\rightarrowfill}}
       \limits_{#1}}}                           

\def\overlay#1#2{\ifmmode%
\setbox0=\hbox{$#1$}%
\setbox1=\hbox to\wd0{\hss$#2$\hss}\else%
\setbox0=\hbox{#1}%
\setbox1=\hbox to\wd0{\hss#2\hss}\fi%
#1\hskip-\wd0\box1 }

\def\pmb#1{\leavevmode\setbox0=\hbox{#1}%
\kern-.02em\copy0\kern-\wd0
\kern.04em\copy0\kern-\wd0
\kern-.02em\raise.04em\box0 }

\def\vereq#1#2{\lower3pt\vbox{\baselineskip1.5pt \lineskip1.5pt
\ialign{$\m@th#1\hfill##\hfil$\crcr#2\crcr\sim\crcr}}}

\def\tensor#1{\protect\@ontopof{#1}{\leftrightarrow}{1.15}\mathord{\box2}}
\def\overstar#1{\protect\@ontopof{#1}{\ast}{1.15}\mathord{\box2}}
\def\overdots#1{\protect\@ontopof{#1}{\cdots}{1.0}\mathord{\box2}}
\def\overcirc#1{\protect\@ontopof{#1}{\circ}{1.2}\mathord{\box2}}
\def\loarrow#1{\protect\@ontopof{#1}{\leftarrow}{1.15}\mathord{\box2}}
\def\roarrow#1{\protect\@ontopof{#1}{\rightarrow}{1.15}\mathord{\box2}}

\def\@ontopof#1#2#3{%
{\mathchoice
{\@@ontopof{#1}{#2}{#3}\displaystyle\scriptstyle}%
{\@@ontopof{#1}{#2}{#3}\textstyle\scriptstyle}%
{\@@ontopof{#1}{#2}{#3}\scriptstyle\scriptscriptstyle}%
{\@@ontopof{#1}{#2}{#3}\scriptscriptstyle\scriptscriptstyle}%
}%
}

\def\@@ontopof#1#2#3#4#5{%
\setbox0=\hbox{$#4#1$}%
\setbox1=\hbox{$#5#2$}%
\setbox2=\hbox{}\ht2=\ht0 \dp2=\dp0 %
\ifdim\wd0>\wd1 %
\setbox1=\hbox to\wd0{\hss\box1\hss}%
\mathord{\rlap{\raise#3\ht0\box1}\box0}%
\else   %
\setbox1=\hbox to.9\wd1{\hss\box1\hss}%
\setbox0=\hbox to\wd1{\hss$#4\relax#1$\hss}%
\mathord{\rlap{\copy0}\raise#3\ht0\box1}%
\fi
}%

\def\lambdabar{\protect\@lambdabar}
\def\@lambdabar{%
\relax
\bgroup
\def\@tempa{\hbox{\raise.73\ht0
\hbox to0pt{\kern.25\wd0\vrule width.5\wd0
height.1pt depth.1pt\hss}\box0}}%
\mathchoice{\setbox0\hbox{$\displaystyle\lambda$}\@tempa}%
{\setbox0\hbox{$\textstyle\lambda$}\@tempa}%
{\setbox0\hbox{$\scriptstyle\lambda$}\@tempa}%
{\setbox0\hbox{$\scriptscriptstyle\lambda$}\@tempa}%
\egroup
}

\def\corresponds{{\lower.2ex\hbox{=}}{\rm\kern-.75em^\triangle}}
\def\succsim{\succ\kern-.9em_\sim\kern.3em}
\def\precsim{\prec\kern-1em_\sim\kern.3em}
\def\slantfrac#1#2{\kern1em^{#1}\kern-.3em/\kern-.1em_{#2}}

\def\alpha{{\Greekmath 010B}}%
\def\beta{{\Greekmath 010C}}%
\def\gamma{{\Greekmath 010D}}%
\def\delta{{\Greekmath 010E}}%
\def\epsilon{{\Greekmath 010F}}%
\def\zeta{{\Greekmath 0110}}%
\def\eta{{\Greekmath 0111}}%
\def\theta{{\Greekmath 0112}}%
\def\iota{{\Greekmath 0113}}%
\def\kappa{{\Greekmath 0114}}%
\def\lambda{{\Greekmath 0115}}%
\def\mu{{\Greekmath 0116}}%
\def\nu{{\Greekmath 0117}}%
\def\xi{{\Greekmath 0118}}%
\def\pi{{\Greekmath 0119}}%
\def\rho{{\Greekmath 011A}}%
\def\sigma{{\Greekmath 011B}}%
\def\tau{{\Greekmath 011C}}%
\def\upsilon{{\Greekmath 011D}}%
\def\phi{{\Greekmath 011E}}%
\def\chi{{\Greekmath 011F}}%
\def\psi{{\Greekmath 0120}}%
\def\omega{{\Greekmath 0121}}%
\def\varepsilon{{\Greekmath 0122}}%
\def\vartheta{{\Greekmath 0123}}%
\def\varpi{{\Greekmath 0124}}%
\def\varrho{{\Greekmath 0125}}%
\def\varsigma{{\Greekmath 0126}}%
\def\varphi{{\Greekmath 0127}}%

\def\nabla{{\Greekmath 0272}}
\def\FindBoldGroup{%
   {\setbox0=\hbox{$\mathbf{x\global\edef\theboldgroup{\the\mathgroup}}$}}%
}

\def\Greekmath#1#2#3#4{%
    \if@compatibility
        \ifnum\mathgroup=\symbold
           \mathchoice{\mbox{\boldmath$\displaystyle\mathchar"#1#2#3#4$}}%
                      {\mbox{\boldmath$\textstyle\mathchar"#1#2#3#4$}}%
                      {\mbox{\boldmath$\scriptstyle\mathchar"#1#2#3#4$}}%
                      {\mbox{\boldmath$\scriptscriptstyle\mathchar"#1#2#3#4$}}%
        \else
           \mathchar"#1#2#3#4%
        \fi 
    \else 
        \FindBoldGroup
        \ifnum\mathgroup=\theboldgroup 
           \mathchoice{\mbox{\boldmath$\displaystyle\mathchar"#1#2#3#4$}}%
                      {\mbox{\boldmath$\textstyle\mathchar"#1#2#3#4$}}%
                      {\mbox{\boldmath$\scriptstyle\mathchar"#1#2#3#4$}}%
                      {\mbox{\boldmath$\scriptscriptstyle\mathchar"#1#2#3#4$}}%
        \else
           \mathchar"#1#2#3#4%
        \fi     	    
	  \fi}

\newif\ifGreekBold  \GreekBoldfalse
\let\SAVEPBF=\pbf
\def\pbf{\GreekBoldtrue\SAVEPBF}%

\makeatother

\begin{document}
\begin{center}
{\Large\textbf{{}{}{}{}Violations of the Weak Energy Condition
for Lentz Warp Drives}}{\Large\par}
\par\end{center}

\begin{center}
 
\par\end{center}

\begin{center}
\medskip{}
 Bill Celmaster \\
 email: \textsl{bill.celmaster@verizon.net} \\
 \medskip{}
 Steve Rubin\\
 email: \textsl{slr42349@gmail.com} \\
\par\end{center}

\begin{center}
\medskip{}
 (\today) 
\par\end{center}
\begin{abstract}
Warp drive spacetimes capable of superluminal transportation, were
first introduced in 1994 by Miguel Alcubierre and then generalized
by others. These spacetimes violated the Weak Energy Condition (WEC).
Lentz proposed a new type of warp drive in 2020. It was claimed that
this warp spacetime has non-negative energy density and can therefore
be sourced by a classical plasma. We demonstrate that Lentz's claim
is incorrect. We begin with a direct calculation of the energy-momentum
tensor of Lentz's warp drive in a Eulerian reference frame, and show
that there are spacetime regions where the energy density is negative.
We then examine the theoretical basis of Lentz's investigation and
identify several derivation errors. The derivation errors can be somewhat
ameliorated with a modified version of Lentz's geometry, which more
closely respects the equalities and inequalities discussed by Lentz.
Even so, we show that the modified geometry still violates the WEC
even in the Eulerian reference frame. We conclude with a brief discussion
of no-go theorems and their relationship to geometries of the kind
proposed by Lentz. 
\end{abstract}

\section{Introduction}

In 1994, Miguel Alcubierre \cite{Alcubierre} described a kind of
spacetime that contained a region he called a ``warp drive''. Warp
drives have several fascinating properties. For example, an observer
whose time line coincides with the center of the warp drive may have
the same clock reading as observers outside the warp drive, that is,
no time dilation, but is traveling at superluminal speed relative
to those observers. Alcubierre then proceeded to show that these warp
drive spacetimes have energy-momentum tensors that violate the weak
energy condition\footnote{Alcubierre notes that other energy conditions are also violated but
in this note, we limit discussion to the weak energy condition.} (WEC), 
\begin{equation}
T_{\mu\nu}t^{\mu}t^{\nu}\ge0,\label{Eq:WEC}
\end{equation}
where $T$ is the energy-momentum tensor, and $t$ is any timelike
vector. It is generally believed that all classical physics theories
have energy-momentum tensors satisfying the WEC and, in particular,
that this holds for any classical plasma. If so, then the Alcubierre
warp drive is either impossible or a consequence of some non-classical
(exotic) theory. Quantum field theories are known to violate the WEC,
but to date, no quantum warp drives have been described (see Barcelo
et al. \cite{Barcelo} and references therein).

Subsequent to the original Alcubierre paper, considerable research
has explored alternate warp drives \cite{Abellan, Santos-Pereira, Santos-Pereira:2021, Natario,  santospereira2025warpdrivesuperluminaltravel, Chargeddussolutions,  Fuchs_2024, Helmerich_2024, Abellan_2024, Bolivar_2025,   VanDenBroeck1999b, PfenningFord1997, Coule1998, GonzalezDiaz2000} and there is even work   done towards the engineering  of warp drives \cite{White2003, White2021}. Over time, various theorems have
been proven showing that for very general classes of warp drive, the
Weak Energy Condition (WEC) must be violated \cite{Natario, Low_1999, Santiago}. 
Moreover, superluminal motion is often associated with time travel (i.e., closed timelike curves).  Although a variety of  proposals -- for general geometries -- have been made for ways in which time travel could possibly avoid paradoxes (see, for example, the review by Shoshany \cite{shoshany1}), it is commonly believed that physical theories should prohibit time travel.  This principle has been named by Hawking \cite{Hawking1992} as ``the chronology protection conjecture".  If a superluminal-capable warp drive were found to comply with the WEC, then we would still want to know whether it violates the chronology protection conjecture.  The question of warp drives and time-travel has been the subject of considerable research \cite{ Olum1998, Everett1996, Visser1993, Visser1994, Visser2003, FriedmanHiguchi2006, Liberati2016, Shoshany}.  For a large category of warp drives including those to be described in what follows, Shoshany \cite{Shoshany} has proven that superluminal travel would indeed permit time travel under certain well-defined circumstances.    

In 2020, Erik Lentz \cite{Lentz,Lentz:2021} proposed a category of
warp drives whose geometries are given in the ADM \cite{Arnowitt}
formalism by specifying shift-vectors (as defined in, for example,
Gourgoulhon \cite{gourgoulhon}) obtained as gradients of potentials
satisfying\textcolor{green}{{} }a certain hyperbolic differential
equation but with various source functions. Lentz then developed a
number of relationships that he claimed are satisfied by all such
warp drives, and from those relationships, he found a non-negativity
condition that, if satisfied, would lead to a non-negative energy
density as measured by an Eulerian observer.
Finally, Lentz proposed a specific source function intended to satisfy
the non-negativity condition, which should therefore lead to a warp
drive geometry that Lentz claimed to satisfy the WEC.  Lentz's paper did not address the issue of time travel, but Shoshany \cite{Shoshany} has proven that if Lentz's warp drive could be built, then it could be configured for time travel.

Lentz's interesting work has received a considerable amount of attention
in the refereed physics literature \cite{Abellan,Fuchs_2024, Helmerich_2024,Santos-Pereira:2021,Chargeddussolutions,Fell,Shoshany,Frasca:2020jzp, Carneiro_2022, Huey_2024,Bolivar_2025, Clough_2024}
as well as in the popular press \cite{Pettit, Borunda, Gast, Skuse,NEUKART2025101676, Houran,Koberlein,Wagh}. Nevertheless, there is a disturbing question of  why Lentz's warp drive appears to violate the previously-mentioned WEC-violation
theorem proven by Santiago et al. \cite{Santiago}.
In fact, Santiago et al. proved an even stronger statement -- namely,
that a general class of warp drives with zero vorticity has a sometimes-negative
Eulerian energy density. This appears to be in direct conflict with
the claim made by Lentz, that his warp drive (which has zero vorticity)
has a non-negative Eulerian energy density. Up to this point in time, the situation has remained confusing.  Either the WEC-violation theorem is wrong, or Lentz's warp drive doesn't satisfy the conditions required for the WEC-violation theorem, or  Lentz's conclusions are wrong.  In what follows, we resolve the matter by demonstrating that Lentz's conclusions are wrong.

We will show by direct analysis of its energy-momentum tensor (as computed from the soliton geometry, via Einstein's field equation) that the Lentz warp drive violates the WEC  and, in particular,
has regions of negative Eulerian energy density. Because these conclusions
differ dramatically from those presented by Lentz \cite{Lentz}, we
present the analysis in detail. There are two key aspects of our demonstration.
First, we show by numerical computation that the Lentz geometry violates
the weak-energy condition. Second, we examine the various relationships
that led to the conclusion in Ref. \cite{Lentz}, that the energy
density is non-negative. It turns out that several of those relationships
are incorrect.

In Section \ref{sec:Nat=00003D00003D00003D00003D00003D0000E1rio},
we introduce a class of warp drives described by Nat{\'a}rio, \cite{Natario}
which generalizes Alcubierre's \cite{Alcubierre} original warp drive,
and we provide some of the key geometric quantities to be used in
the subsequent analysis. In Section \ref{sec:Lentz}, we review the
defining characteristics -- such as the differential equation to
be solved by the potential whose gradient is the shift vector --
of Lentz's warp drive geometries. We then examine Lentz's rhomboidal
source (LRS) warp drive and show that its Eulerian energy has regions
of negativity, thus violating the WEC. We point out that this might
be due to the fact that the LRS geometry \uline{does not} in fact
satisfy Lentz's warp-drive defining characteristics. In Section \ref{sec:generalHSVP},
we introduce a class of geometries that (mostly) satisfy Lentz's warp-drive
defining characteristics. We then turn to the question of why Lentz
expected his warp drive to satisfy the WEC. In Ref. \cite{Lentz},
Lentz derived a number of equations and inequalities that lead to
criteria for WEC-compliance. However, the LRS warp drive does not
satisfy these criteria. in Section \ref{sec:generalHSVP} we show
that Lentz made several mistakes in the derivations of the expressions
that led him to his WEC-compliance criteria. In Section \ref{sec:phiMod}
we introduce and examine a modified version (MLRS) of the LRS warp
drive. This modified version is a member of the class of geometries
considered in the previous section, also based on a rhomboidal source
such as Lentz's original proposal. We show that this warp drive violates
the equations, inequalities and properties proposed by Lentz, and
that in the previous section were seen to be generally incorrect.
In Section \ref{sec:SrcRect} we replace the rhomboids by rectangles
to determine whether our conclusions are sensitive to the type of
source function chosen. Again, this geometry violates the WEC, but
the non-zero regions of energy density are highly concentrated near
the vertex points. This example is noteworthy because it significantly
differs from the energy distribution of Alcubierre's (\cite{Alcubierre})
original warp drive. The last topic we explore, in Section \ref{sec:no-go},
is the question of whether no-go theorems should apply to the warp
drives considered in this paper. If so, do those no-go theorems depend
on the conditions that, if relaxed, could violate those theorems and
therefore allow a WEC-compliant warp drive? Here, we briefly address
these possibilities.

\section{Nat{\'a}rio\ warp drives and the Eulerian energy density condition}

\label{sec:Nat=00003D00003D00003D00003D00003D0000E1rio}

We limit our discussion to the spacetimes described by Nat{\'a}rio,\cite{Natario}
which dominate most of the literature on warp drives. These spacetimes
have line elements, 
\begin{equation}
ds^{2}=-(1-N^{i}N_{i})dt^{2}-2N_{i}dx^{i}dt+\sum_{i=1}^{3}dx^{i2},\label{Eq:metric}
\end{equation}
where we have written the line element using the convention of the
ADM formalism\textcolor{black}{\cite{Arnowitt},} where the shift
vector $N_{i}$ is a function of space and time.\footnote{The sign of the shift-vector is a matter of convention. We have adopted
the convention used by Lentz. This choice has no impact on the sign
of the Eulerian energy density.} Spacetime is foliated into flat spacelike hypersurfaces of constant
$t$. Coordinates on the hypersurfaces are indexed using Latin letters
and, because the hypersurfaces are flat, all hypersurface raising
and lowering operators are identity operators. So, for example, $N_{i}=N^{i}$.
For certain values of the shift vectors, the resulting spacetimes
describe warp drives of the type introduced by Alcubierre\cite{Alcubierre}.
We now define some tensors that are important in the analysis.

The normal vector $n$ is defined as 
\begin{equation}
n^{\mu}=(1,N^{i}),\label{eq:normal}
\end{equation}
where we adopt the convention that Greek indices refer to space-time
and as mentioned above, Latin indices refer to the spatial coordinates
of the level-$t$ hypersurfaces. Note that the normal vector is timelike.
If the observers have a 4-velocity $n^{\mu}$, we refer to them as
\textit{Eulerian observers}.

We also introduce the extrinsic curvature tensor for a particular
hypersurface $t=\text{constant}$. It is defined as 
\begin{equation}
K_{ij}=\frac{1}{2}\left(\partial_{i}N_{j}+\partial_{j}N_{i}\right).\label{Eq:extrinsic}
\end{equation}

Recall from Eq. (\ref{Eq:WEC}) that the weak energy condition requires
$T_{\mu\nu}t^{\mu}t^{\nu}\ge0$ for all time-like vectors $t$. A
fortiori, $T_{\mu\nu}n^{\mu}n^{\nu}\ge0$. This quantity, $T_{\mu\nu}n^{\mu}n^{\nu}$
is known as the Eulerian energy density density $E$ (i.e. the energy
density as measured by a Eulerian observer), and the inequality $T_{\mu\nu}n^{\mu}n^{\nu}\ge0$
is ``the Eulerian energy density condition''.

Clearly, to prove that the WEC inequality is obeyed, we must demonstrate
the inequality for \textbf{all} timelike vectors, and it is insufficient
to establish the inequality only for the Eulerian energy density.
However, to prove that the WEC is violated, it suffices to demonstrate
the violation of the Eulerian energy density condition. This is what
we will show in the sections which follow.\footnote{Note that Lentz \cite{Lentz} claims to show that his warp drive has
a non-negative Eulerian energy density. Had his conclusion been correct,
it still wouldn't have established the validity of the WEC, since
other time-like vectors would have needed to be considered.}

One of the virtues of the ADM formalism is that it permits Einstein's
equations to be formulated purely in terms of the extrinsic tensor.
It can be shown (for example, see Gourgoulhon \cite{gourgoulhon})
that 
\begin{equation}
8\pi E=\frac{1}{2}\left(-K_{i}^{j}K_{j}^{i}+K^{2}\right),\label{Eq:Eulerian1}
\end{equation}
where $E$ is the Eulerian energy density.

\section{Lentz's original warp drive }

\label{sec:Lentz}

In his paper \cite{Lentz}, Lentz proposed a warp-drive geometry that
he argued had a non-negative energy density. His non-negativity argument
depends on constructing the geometry from a potential function $\phi$
that meets several criteria, including being a solution to a differential
equation -- Eq. (\ref{Eq:de}) below -- with a rhomboidal source
function $\rho_{\text{rhomb}}$ -- Eq. (\ref{Eq:srcRhom}) below.
Lentz then presents a potential function $\phi_{L}$ that he claims
satisfies the solution criteria. He next claims to show, by numerical
computation and display of the Eulerian energy density, that his geometry
satisfies the WEC. We demonstrated that Lentz made a mistake in his
computation of the energy density, and that in fact, it is not positive
and violates the WEC. Lentz's mistake arises, in part, from the fact
that $\phi_{L}$ is \textbf{not} a solution to Eq. (\ref{Eq:de}),
and not even approximately. Because Lentz's WEC compliance (non-negativity)
argument was supposedly derived from the solutions of Eq. (\ref{Eq:de}),
it is not surprising that an incorrect solution to that equation leads
to a WEC violation.

In Section \ref{sec:phiMod}, we propose a modified version of $\phi_{L}$
that closely resembles what Lentz presumably had in mind. The function
$\phi_{\text{rh}}$ is a solution to Eq. (\ref{Eq:de}) almost everywhere
(however, we were unable to find a solution for all values of the
coordinates).

\subsection{The warp drive hyperbolic differential equation}

\label{subsec:de}

Lentz \cite{Lentz} defined a potential function $\tilde{\phi}$ as
a solution to the hyperbolic differential equation (Eq.(15) in Ref.
\cite{Lentz}) 
\begin{equation}
\begin{aligned}\partial_{x}^{2}\tilde{\phi}+\partial_{y}^{2}\tilde{\phi}-\frac{2}{v_{h}^{2}}\partial_{z}^{2}\tilde{\phi} & =\tilde{\rho},\end{aligned}
\label{Eq:de}
\end{equation}
where $\tilde{\rho}$ denotes the specified source function. For later
reference, we refer to Eq. (\ref{Eq:de}) as ``Lentz's differential
equation". The warp geometry is obtained by substituting the shift
vector 
\begin{equation}
\textbf{N}=\nabla\tilde{\phi}\label{Eq:grad}
\end{equation}
into the metric defined in Eq. (\ref{Eq:metric}).\textcolor{green}{{}
}Lentz then imposes additional conditions on $\tilde{\phi}$ for which
it is a unique solution. One of these conditions is on the boundary
$\tilde{\phi}\to0$ as $z\to-\infty$.\textcolor{green}{{} }The second
condition is a constraint that reduces the dimensionality of the function
domains; namely, $\tilde{\rho}$ can be written as\footnote{Although Lentz doesn't explicitly state the condition $\rho(s(x,y),z)=\rho(-s(x,y),z)$,
it can be inferred from his Fig. (1).} $\tilde{\rho}(x,y,z)=\rho(s(x,y),z)=\rho(-s(x,y),z)$ and $\tilde{\phi}$
can be written as $\tilde{\phi}(x,y,z)=\hat{\phi}(s(x,y),z)$,\textcolor{green}{{}
}where $s(x,y)=|x|+|y|$, which implies that the solution $\tilde{\phi}$
should be symmetric under the exchange of $x$ and $y$. Finally,
Lentz makes $\tilde{\rho}$ and $\tilde{\phi}$ implicit functions
of time, by setting $(x,y,z)\to(x,y,z-vt)$. We refer to $\textbf{v}=(0,0,v)$
as the ``velocity of the warp drive."

\subsection{Lentz's rhomboidal source}

\label{subsec:SrcRhom}

The remaining step in specifying the Lentz metric is to select the
source. At this stage, we must point out that Lentz does not provide
an analytical definition of his source. Therefore, we constructed
a source $\rho_{\text{rhomb}}$, that resembles Lentz' as closely
as possible based on the information provided. Lentz chooses his source
as a function of $|x|+|y|$ and $z$.\textcolor{blue}{{} }He also
describes it as consisting of rhomboids in the $y=0$ plane, and then
provides a picture, Fig. (1) in ref. \cite{Lentz}.\textcolor{green}{{}
}One of our challenges was to find an analytical form for $\rho_{\text{rhomb}}$
which results in a picture similar to that of Lentz.\footnote{To further verify that we have used a source virtually identical to
Lentz's we note below that the $x$ and $z$ components of the shift
vector resemble their counterparts plotted in ref.\cite{Lentz}. The
sign difference turns out to be due the fact that Lentz's figure of
$N_{x}$ is inconsistent with his Eq. (20), in that the signs are
reversed.}

The source function must also be such that the resultant shift vector
$\textbf{N}$ has the following properties. 
\begin{enumerate}
\item The $z$-component $N_{z}$ of the shift vector is non-zero and level
over a finite region about the origin (which Lentz refers to as the
``central region''), 
\item The $x$ and $y$-components of the shift vector are zero over the
same region. 
\end{enumerate}
A further requirement is that Eulerian observers in the central region
move synchronously with the warp drive. This requirement is satisfied
by including in the definition of $\rho_{\text{rhomb}}$ an overall
scale factor $W=v/N_{z}(0,0,0)$, where $v$ is the $z$-component
of the warp drive velocity (as described in Section \ref{subsec:de}).
We note that this is only approximately true for the rhomboidal source
described below, because properties 1 and 2 above are only approximately
true. However, in Section \ref{sec:SrcRect}, we provide an alternative
source such that these conditions hold exactly over a broader region.
In the body of this paper, we will not further discuss time-dependence
or velocity, but in Appendix \ref{sec:F-superlumin} we show how the
class of observers just described, experience travel, including superluminal
travel, without time dilation.

Here is the analytic form that we used for the rhomboidal-source function.
\begin{equation}
\begin{aligned}\rho_{\text{rhomb}}(s,z)=\tilde{\rho}_{\text{rhomb}}(x,y,z)=W\sum_{i=1}^{4}\alpha_{i}\Theta\left(p_{i}(s,z)\right)g_{i}(s),\end{aligned}
\label{Eq:srcRhom}
\end{equation}
where $\Theta$ is the Heaviside function and 
\begin{equation}
p_{i}(s,z)=L_{x}-\frac{L_{x}}{L_{z}}|z-\xi_{i}|-\left||s|-\beta_{i}\right|,\label{Eq:shapeRhom}
\end{equation}
and 
\begin{equation}
g_{i}(s)=Q_{\text{min}}+\left(Q_{\text{max}}-Q_{\text{min}}\right)(|s|-\beta_{i})^{2},\label{Eq:chgPflRhom}
\end{equation}
where:\textcolor{green}{{} } 
\begin{itemize}
\item $W$ is the overall scale factor, described above. 
\item $s=|x|+|y|$ 
\item $L_{x}$ is the semi-length of the rhomboids along the $x$-direction. 
\item $L_{z}$ is the semi-length of the rhomboids along the $z$-direction. 
\item $\alpha_{i}$ are weights that may be negative or positive 
\item $\beta_{i}$ are the $s$-coordinates of the rhomboid centroids. 
\item $\xi_{i}$ are the $z$-coordinates of the rhomboid centroids. 
\item $Q_{\text{max}}$ and $Q_{\text{min}}$ are the maximum and minimum
charge, respectively. 
\end{itemize}
In addition, this source configuration satisfies the following constraints:\textcolor{green}{{}
} 
\begin{itemize}
\item Rhomboids are elongated along the $x$-direction and narrow along
the $z$-direction, as shown in Fig. (1) in ref.\cite{Lentz}, i.e.
$L_{x}\gg L_{z}$. 
\item Rhomboids are equally spaced along the $z$-direction, that is $\triangle\xi$
is a constant, where $\triangle\xi$ is the difference in $\xi_{i}$
between any rhomboid and its nearest neighbor along the $z$-direction.
For any such pair of rhomboids, the corresponding difference $\triangle\beta$
is also constant. This implies that the line joining the centroids
of any such pair of rhomboids creates a constant angle $\theta=\arctan\left(\pm\triangle\beta/\triangle\xi\right)$
with the $z$-axis.\\
 \\
 Note: It will be seen that, in order to guarantee that the potential
$\phi$ and the shift vector $N_{i}$ are bounded (at $y=0$ in Lentz's
original ansatz and for all $y$ in the corrected ansatz presented
in Section \ref{sec:generalHSVP}), it is necessary that $\theta=\arctan\left(\pm1/v_{h}\right)$.
Therefore, as a consequence of setting $v_{h}=1$, we must have $\theta=\pm45^{o}$
or equivalently, $\triangle\beta=\pm\triangle\xi$. The choices for
the rhomboid centroids $\left(\beta_{i},\xi_{i}\right)$, shown in
the table below, reflect this more stringent constraint. 
\item Rhomboids do not overlap, i.e. $\triangle\xi\geq L_{z}$ and, unless
$\beta_{i}=0$, $\left|\beta_{i}\right|\geq L_{x}$ 
\item The total integrated charge of the system is zero, implying that $\sum_{i=1}^{4}\alpha_{i}=0$,
where $\alpha_{i}\neq0$ for all $i$. 
\end{itemize}
The source configuration is such that a rotation of $180^{\circ}$
about the $z$-axis leaves the rhomboids unchanged, that is unless
$\beta_{i}=0$, the $\beta_{i}$ come in pairs such that, for each
rhomboid with parameters $\beta_{i},\alpha_{i},\xi_{i}$, there is
another rhomboid with parameters $-\beta_{i},\alpha_{i},\xi_{i}$.
Thus, as a result of the preceding constraints, Eqs. (\ref{Eq:shapeRhom})
and (\ref{Eq:chgPflRhom}) imply that the complete source configuration
comprises seven rhomboids.

For the computational results that follow (see Appendices \ref{sec:C-evalIntegrals},
\ref{sec:D-derivatives}, \ref{sec:E-graphics}), we used the following
values for the parameters, which satisfy the definitions and constraints
discussed above.

\begin{tabular}{|c|c|}
\hline 
Parameter  & Value(s)\tabularnewline
\hline 
\hline 
\begin{tabular}{|c|}
\hline 
$i$\tabularnewline
\hline 
\hline 
$\beta_{i}$\tabularnewline
\hline 
$\xi_{i}$\tabularnewline
\hline 
$\alpha_{i}$\tabularnewline
\hline 
\end{tabular} & %
\begin{tabular}{|c|c|c|c|}
\hline 
$1$  & $2$  & $3$  & $4$ \tabularnewline
\hline 
\hline 
$1$  & $2$  & $1$  & $0$ \tabularnewline
\hline 
$-1$  & $0$  & $1$  & $2$ \tabularnewline
\hline 
$25$  & $-25$  & $-25$  & $50$ \tabularnewline
\hline 
\end{tabular}\tabularnewline
\hline 
$L_{x}$  & $1$\tabularnewline
\hline 
$L_{z}$  & $1/8$\tabularnewline
\hline 
$W$  & $v/N_{z}(0,0,0),\:v=1$\tabularnewline
\hline 
$Q_{\text{max}},Q_{\text{min}}$  & $1,$1/4\tabularnewline
\hline 
\end{tabular}

\begin{figure}[H]
\centering \includegraphics[scale=0.72]{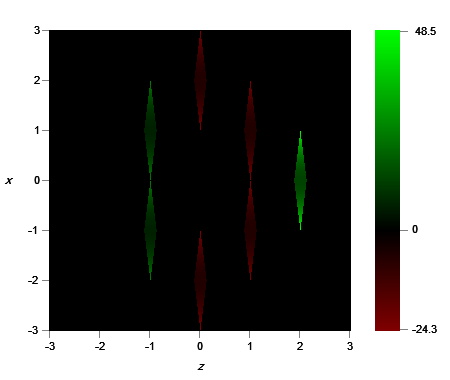} \caption{Unscaled rhomboidal source $\tilde{\rho}_{\text{rhomb}}N_{z}(0,0,0)$
for $y=0$}
\label{Fig:Src_rhom} 
\end{figure}

The rhomboid source\textcolor{green}{, }as described in Eqs. (\ref{Eq:srcRhom}),
(\ref{Eq:shapeRhom}) and (\ref{Eq:chgPflRhom}), is displayed in
Fig.(\ref{Fig:Src_rhom})\textcolor{green}{, }except that it was not
scaled by $W$. That is, the figure shows $\tilde{\rho}_{\text{rhomb}}/W$,
which is equivalent to $\tilde{\rho}_{\text{rhomb}}N_{z}(0,0,0)$,
as $v=1$. We do this so that the figure may be more easily compared
to Lentz's Fig. 1.

As mentioned earlier, when we compare Fig. (\ref{Fig:Src_rhom}) to
Fig. (1) in ref. \cite{Lentz}, we can see that the two source functions
appear very close to one another. We also varied some of the parameters
of $\rho_{\text{rhomb}}$ to see how these parameters would affect
our conclusions, and found no qualitative differences.

\subsection{Lentz's potential, $\phi_{L}$}

\label{subsec:phiLentz}

In this section, we begin with Lentz's function $\phi_{L}$ and carefully
use it to build a warp geometry. Subsequently, we compute the Eulerian
energy density. Despite using the same geometry as Lentz, we do not
obtain the non-negative result that he shows in \cite{Lentz}. We
trace this discrepancy in part to a sign error (see Appendix \ref{sec:A-MotivPhi})
that Lentz made in one of the terms that he uses in his calculation,
which is ultimately caused by the fact that $\phi_{L}$ is not a solution
of Lentz's differential equation, Eq. (\ref{Eq:de}). Moreover, had
Lentz displayed his result at greater $\gamma$-correction (see Appendix
\ref{sec:E-graphics}), his `positive' result (using the wrong sign)
would have revealed some regions of negativity.

Lentz \cite{Lentz} defines the potential $\phi_{L}(x,y,z)$ by setting
\begin{equation}
\begin{aligned}\phi_{L}(x,y,z(t)) & =\frac{1}{4v_{h}}\int_{-\infty}^{\infty}dx'dz'\Theta\left(z(t)-z'-\frac{1}{v_{h}}|x-x'|\right)\rho_{\text{rhomb}}(|x'|+|y|,z'(t)),\end{aligned}
\label{Eq:phiLentz}
\end{equation}
where $z(t)=z-vt$ and $v$ is the speed of the warp drive. This is
equivalent to Eq. (18) in ref. \cite{Lentz}. In this section, we
set $v_{h}=1$. Lentz claimed that $\phi_{L}$ is a solution to the
differential equation in Eq. (\ref{Eq:de}) with the source in Eq.
(\ref{Eq:srcRhom}) and under the conditions specified in Subsection
(\ref{Fig:Nz_rhom}). As shown in Appendix \ref{sec:A-MotivPhi},
this claim is false. \textbf{$\phi_{L}$ is not a solution.} As an
example of one of the ways in which the claim is false, note that
$\phi_{L}$ is not symmetric under the exchange of $x$ and $y$,
and therefore violates one of Lentz's solution conditions.

We computed the Eulerian energy, by first computing the shift vector.
We substitute $\rho_{\text{rhomb}}$ into the expression for $\phi_{L}$
given by Eq. (\ref{Eq:phiLentz}) and then computing the shift vector
by applying Eq. (\ref{Eq:grad}). Details of the calculations are
provided in Appendix \ref{sec:C-evalIntegrals}.

The rhomboidal-source shift-vector component $N_{z}^{L}$ is displayed
in Fig.(\ref{Fig:Nz_rhom}).

\begin{figure}[H]
\centering \includegraphics[scale=0.72]{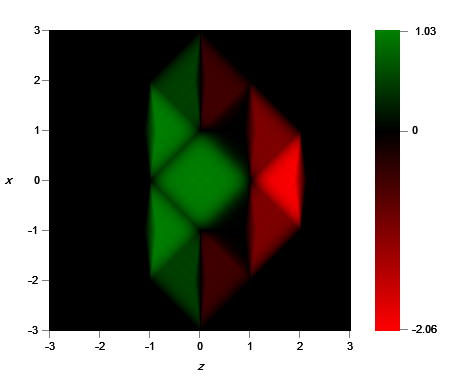} \caption{ $N_{z}^{L}$ for the rhomboidal source, evaluated on the slice y=0.}
\label{Fig:Nz_rhom} 
\end{figure}

\begin{figure}[H]
\centering \includegraphics[scale=0.72]{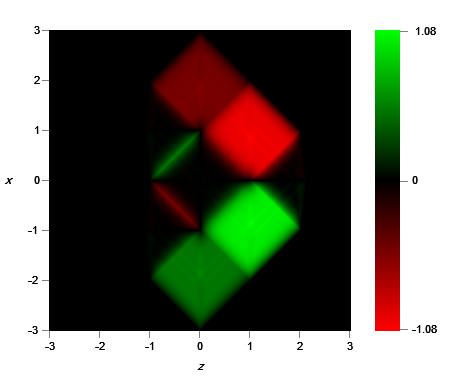} \caption{$N_{x}^{L}$ for the rhomboidal source, evaluated on the slice $y=0$.}
\label{Fig:Nx_rhom} 
\end{figure}

This figure can be compared with the left-hand side of Fig. (2) in
Lentz \cite{Lentz}:\footnote{Our definition of $\mathbf{N}$ already includes the factor $v/N_{z}(0,0,0)$
that Lentz shows explicitly, where $v=1$. (This note also applies
to Fig. (\ref{Fig:Nx_rhom}).)} We see that the plots are very similar. We also compared $N_{x}^{L}$
with Lentz's figure for $N_{x}^{L}$ (right-hand side of Fig. (2)
of Lentz \cite{Lentz}). The results are shown in Fig. (\ref{Fig:Nx_rhom}).

It turns out that Lentz's figure of $N_{x}^{L}$ is inconsistent with
his Eq. (20), in that the signs are reversed. Nevertheless, it is
structurally the same as our result and thus provides further evidence
that our source function must be close to that of Lentz.\footnote{As far as we can tell, none of Lentz's other results are sensitive
to this sign-inconsistency.}

\begin{figure}[H]
\centering \includegraphics[scale=0.72]{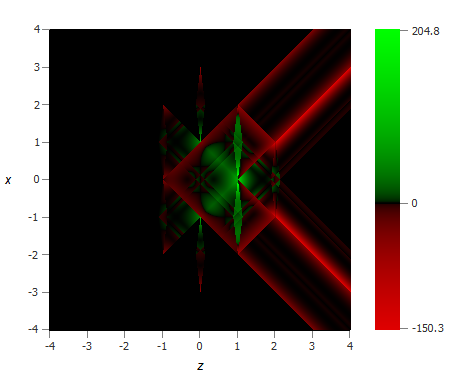} \caption{Rhomboidal-source Eulerian energy density at $y=1.0\times10^{-6}.$}
$\left(\gamma=0.43\right)$ \label{Fig:E_L_rhom} 
\end{figure}

Finally, we computed the Eulerian energy density $E_{L}$ using Eqs.
(\ref{Eq:Eulerian1}) and (\ref{Eq:extrinsic}) respectively. This
computation was performed using the analytical methods described in
Appendix \ref{sec:C-evalIntegrals} for the first derivatives and
in Appendix \ref{sec:D-derivatives} for the second derivatives. Note
that $E_{L}$ is ambiguous (not uniquely defined) for certain sets
of points including the plane $y=0$, owing to discontinuities in
$N_{y}$ (See Appendix \ref{sec:G-yDep}). We address the issue of
discontinuities more fully in the next section and Appendix \ref{sec:D-derivatives}.
In addition, as shown in Fig. (\ref{Fig:E_L_rhom}), we avoided the
ambiguity on the $y=0$ plane by displaying $E_{L}$ for $y=1.0\times10^{-6}$.

It can be observed from Fig. (\ref{Fig:E_L_rhom}) that there are
regions of both positive and negative energy densities; therefore,
this warp drive violates the WEC.

Note that in Fig. (3) of Lentz \cite{Lentz}, the Eulerian energy
density appears very different, and in fact looks non-negative. Since
this figure is offered as one of the main conclusions of Lentz's paper
-- namely, that the Lentz geometry is WEC compliant -- it is important
to re-emphasize that Lentz's figure is \textbf{ not} correctly derived
from his geometry, and to give a detailed explanation of why. We believe
that there are two significant errors that contribute to Lentz's incorrect
demonstration of energy positivity. 
\begin{itemize}
\item Lentz uses his Eq. (17) to obtain his Fig. (3), an assumption we have
made based on the fact that we can reproduce the apparent positive
energy density shown in his Fig. (3) using Eq. (17) to compute the
Eulerian energy density, and then displaying the result by setting
$\gamma=1$ (no gamma correction). However, Lentz's Eq. (17) is not
the correct equation for the energy density of the geometry. The equation
is derived by assuming that $\phi_{L}$ is a solution of Lentz's differential
equation -- Eq. (\ref{Eq:de}). As previously mentioned, this assumption
is false. In particular, the illusion of a positive energy density
is largely the result of an overall sign error in the expression for
$\phi_{L}$, as mentioned at the beginning of this section and demonstrated
in Appendix \ref{sec:A-MotivPhi}. 
\item Although Fig. (3) of Lentz \cite{Lentz} shows a positive energy density
everywhere, this is an artifact of the graphical display, assuming
that Lentz employed his Eq. (17), as previously discussed. When we
display that result by setting $\gamma$ to low values (greater correction),
we can clearly see regions of negativity that are almost invisible
for $\gamma=1$. 
\end{itemize}
There is one more observation regarding our display -- Fig. (\ref{Fig:E_L_rhom})
for Eulerian energy density. We can see that the energy density does
not appear confined close to the region where the source is non-zero.
In fact, there are two $45^{\circ}$ bands extending to large positive
values of $z$ which presumably extend to infinity. (To make this
more obvious, we show a broader region of space than in other figures.)
Similar results were observed for the other values of $y$. Consequently,
the energy obtained by integrating the energy density, is infinite.
This fact violates an implicit, and hitherto unmentioned, requirement
of the potential, namely, that it must result in a geometry with a
finite total energy. Why does finite-energy violation occur? In Ref.
\cite{Lentz}, Lentz speaks of how ``sources are organized to terminate
the stray branches of the wave cone". However, because of the asymmetry
between $x$ and $y$, this precise organization becomes increasingly
distorted as $y$ gets further away from zero. For further discussion
on this asymmetry, see Appendix \ref{subsec:A.1}.

\section{Properties of hyperbolic potential (HP) warp drives}

\label{sec:generalHSVP}

We turn now to Lentz's non-negativity ``derivations''\footnote{We have put quotation marks around the word ``derivations", because
-- as we shall see -- some of the ``derivations" are incorrect.} for a class of warp geometries based on shift vectors $\textbf{N}=\nabla\phi$
where 
\begin{equation}
\phi(x,y,z)=-\frac{1}{4}\int_{-\infty}^{+\infty}\int_{-\infty}^{+\infty}ds'dz'\Theta(z-z'-|s(x,y)-s'|)\rho(s',z'),\label{Eq:phiRSVP}
\end{equation}
and $s(x,y)=|x|+|y|$, $\rho(s(x,y),z)=\tilde{\rho}(x,y,z)$ and $\rho$
is the source function to be specified. We refer to such potentials
$\phi$ as ``hyperbolic potentials'' (HP). We will examine the properties
of HP geometries, including the number of relationships from which
Lentz incorrectly concluded that certain HP geometries must satisfy
the WEC. We will see that these claims are incorrect.

\subsection{HP warp drives almost satisfy the Lentz DE}

\label{subsec:DEsolution}

By taking appropriate the derivatives of $\phi$, we can show that
it satisfies Lentz's differential equation Eq. (\ref{Eq:de}) almost
everywhere.\footnote{\label{note1}We use the (non-standard) phrase \textit{almost everywhere}
to mean ``everywhere except for delta functions.''} The following identities prove useful.\footnote{We have used the distribution equality $\partial_{x}|x|=\mathrm{sgn}(x)$.
It is also common to write this equality as $\partial_{x}|x|=\frac{x}{|x|}$.
Since these expressions are discontinuous at $x=0$, there is an ambiguity
(explained earlier in the text) about how to extend the expressions
to $x=0$. Considered as distributions, all finite definitions will
be equivalent, so we won't offer a definition for $x=0$ other than
to require finiteness.} 
\begin{equation}
\begin{aligned}\partial_{x}f(s(x,y),z) & =\partial_{s}f(s,z)\mathrm{sgn}(x),\\
\partial_{y}f(s(x,y),z) & =\partial_{s}f(s,z)\mathrm{sgn}(y),\\
\partial_{s}\phi & =\frac{1}{4}\int_{-\infty}^{\infty}ds'\mathrm{sgn}(s-s')\rho(s',z-|s-s'|),
\end{aligned}
\label{eq:identities}
\end{equation}
for any differentiable function $f$. After some manipulations (some
of which are presented in Appendix \ref{sec:A-MotivPhi}), we obtain
\begin{equation}
\begin{aligned}\partial_{x}^{2}\phi+\partial_{y}^{2}\phi-2\partial_{z}^{2}\phi= & \rho(s,z)+S(x,y,z)\end{aligned}
\label{Eq:DE}
\end{equation}
where 
\begin{equation}
S(x,y,z)=2\left(\delta(x)+\delta(y)\right)\partial_{s}\phi=\frac{1}{2}\left(\delta(x)+\delta(y)\right)\int_{-\infty}^{+\infty}ds'\mathrm{sgn}(s-s')\rho(s',z-|s-s'|).\label{Eq:Sxy}
\end{equation}
This demonstrates that $\phi$ is a solution to Lentz's differential
equation, except when $x=0$ or $y=0$. These boundary terms are physically
consequential. We were unable to find a closed-form potential $\tilde{\phi}$
that solved Lentz's differential equation.

\subsection{ Energy density of HP warp drives}

\label{subsec:EnergyDensity}

In this section, we derive the following: if $\phi$ is a solution
to Eq. (\ref{Eq:DE}), the energy density $E$ is given by 
\begin{equation}
E=\frac{1}{8\pi}\left[T(x,y,z)+U(x,y,z)\right]\label{Eq: EnergyDensity}
\end{equation}
where 
\begin{equation}
T(x,y,z)=S(x,y,z)\left(\partial_{s}^{2}\phi+\partial_{z}^{2}\phi\right)+4\delta(x)\delta(y)\left(\partial_{s}\phi\right)^{2}.\label{Eq:Txy}
\end{equation}
and 
\begin{equation}
U(x,y,z)=\left(\rho(s,z)+2\partial_{z}^{2}\phi\right)\partial_{z}^{2}\phi-2\left(\partial_{z}\partial_{s}\phi\right)^{2}.\label{Eq:Uxyz}
\end{equation}
This expression for energy density contains the term $U$ which is
similar to Lentz's Eq. (17) but also adds the term $T(x,y,z)$, which
is zero almost everywhere\footref{note1} and is proportional to
delta functions. As we will see later, the delta functions can contribute
significantly to the total energy.

Our derivation of Eqs. (\ref{Eq: EnergyDensity}) - (\ref{Eq:Uxyz}),
starts with the Eulerian energy density, $E$, as given by Eqs. (\ref{Eq:extrinsic})
and (\ref{Eq:Eulerian1} ) as follows:

\[
8\pi E=\frac{1}{2}\left(-K_{i}^{j}K_{j}^{i}+K^{2}\right),
\]
where 
\[
K_{ij}=\frac{1}{2}\left(\partial_{i}N_{j}+\partial_{j}N_{i}\right),
\]
with the shift-vector given by Eq. (\ref{Eq:grad}) 
\[
\textbf{N}=\nabla\phi.
\]

Expanding all the terms, we get 
\begin{equation}
\begin{aligned}E=\frac{1}{8\pi}\left(\partial_{x}^{2}\phi\partial_{y}^{2}\phi+\left(\partial_{x}^{2}\phi+\partial_{y}^{2}\phi\right)\partial_{z}^{2}\phi-\left(\partial_{x}\partial_{y}\phi\right)^{2}-\left(\partial_{x}\partial_{z}\phi\right)^{2}-\left(\partial_{z}\partial_{y}\phi\right)^{2}\right).\end{aligned}
\label{Eq:Eulerian2}
\end{equation}
Then we use Eq. (\ref{Eq:DE}) above to substitute for $\left(\partial_{x}^{2}\phi+\partial_{y}^{2}\phi\right)$.
\begin{equation}
\begin{aligned}E & =\frac{1}{8\pi}\left(\partial_{x}^{2}\phi\partial_{y}^{2}\phi+\left(\rho(s,z)+2\partial_{z}^{2}\phi+S(x,y,z)\right)\partial_{z}^{2}\phi-\left(\partial_{x}\partial_{y}\phi\right)^{2}-\left(\partial_{x}\partial_{z}\phi\right)^{2}-\left(\partial_{z}\partial_{y}\phi\right)^{2}\right).\end{aligned}
\label{Eq:Eulerian3}
\end{equation}

To further modify the above expression for $E$, note that 
\begin{equation}
\begin{aligned}\partial_{x}^{2}\phi & =\partial_{x}\left(\mathrm{sgn}(x)\partial_{s}\phi\right)=2\delta(x)\partial_{s}\phi+\partial_{s}^{2}\phi.\\
\partial_{y}^{2}\phi & =2\delta(y)\partial_{s}\phi+\partial_{s}^{2}\phi.\\
\partial_{x}\partial_{y}\phi & =\mathrm{sgn}(x)\mathrm{sgn}(y)\partial_{s}^{2}\phi.\\
\partial_{z}\partial_{x}\phi & =\mathrm{sgn}(x)\partial_{s}\partial_{z}\phi.\\
\partial_{z}\partial_{y}\phi & =\mathrm{sgn}(y)\partial_{s}\partial_{z}\phi.
\end{aligned}
\end{equation}

Substituting in Eq. (\ref{Eq:Eulerian3}), we obtain Eqs. (\ref{Eq: EnergyDensity}),
(\ref{Eq:Txy}) and (\ref{Eq:Uxyz}).

\subsection{Finiteness of the energy of HP warp drives}

\label{subsec:Finiteness}

One of the requirements for the stress tensor, and therefore for the
potential $\phi$ above, is that the total energy is finite in any
reference frame. This condition is satisfied provided that the source
function $\rho$ in Eq. (\ref{Eq:phiRSVP}) satisfies the following
three conditions: 
\begin{itemize}
\item The linear-cancellation condition

\begin{equation}
\int_{-\infty}^{\infty}ds'\rho(s',\alpha\pm s')=0\label{Eq:cancellation}
\end{equation}
for any constant $\alpha$. 
\item The source-domain boundedness condition 
\begin{equation}
\begin{aligned} & \rho\;\text{has a bounded domain such that}\;\rho(s,z)=0\\
 & \;\text{if}\;|z|>B\;\text{or}\;|s|>B
\end{aligned}
\label{Eq:boundedness}
\end{equation}
for some $B$. 
\item $s(x,y)=|x|+|y|$ 
\end{itemize}
In Appendix \ref{sec:H-Boundedness}, we demonstrate that, when these
conditions hold, the shift vector components have a bounded domain.
That is, there exists a bound $C$, such that $N_{i}(x,y,z)=0$ unless
$|x|,|y|,|z|\le C$. From this, it follows that the stress tensor
has a bounded domain and thus, the integral of $T^{\mu\nu}t_{\mu}t_{\nu}$
for timelike $t$, is finite over any spatial slice.

For Lentz's original warp drive, introduced in the previous section,
the boundedness of the shift vector component $N_{z}^{L}$ is portrayed
at $y=0$ in Fig.(\ref{Fig:Nz_rhom}). Similarly, for $N_{x}$ in
Fig.(\ref{Fig:Nx_rhom}). However, in Lentz's original warp drive,
this boundedness is only true at $y=0$. At the other values of $y$
the shift vector is unbounded. We have observed this implicitly in
Fig.(\ref{Fig:E_L_rhom}), where the Lentz warp-drive energy density
appears unbounded when $y=1.0\times10^{-6}$.

\subsection{The energy non-negativity condition and shift-derivative inequality}

\label{subsec:EulerLentz}

Lentz proposed the inequality, in his Eq. (13), which is a non-negativity
condition for the Eulerian energy density. Lentz's inequality depends
on a shift-derivative inequality $(\partial_{z}^{2}\phi)^{2}\ge(\partial_{z}\partial_{s}\phi)^{2}$,
which is false in many cases including our rhomboidal geometry introduced
in the next section. Moreover, even when this assumption is true,
Lentz's theorem fails when $x=0$ or $y=0$. This failure has serious
consequences for Lentz's non-negativity condition. The remainder of
this section elaborates on this summary.

\subsubsection{The shift-derivative inequality}

The shift-derivative inequality is 
\begin{equation}
(\partial_{z}^{2}\phi)^{2}\ge(\partial_{z}\partial_{s}\phi)^{2}.\label{Eq:ShiftVectorIneq}
\end{equation}
Lentz \cite{Lentz} claims (see the text between his Eqs. (19) and
(20)) that he has proven a version of this and that it can be derived
from Eqs. (19) and (20) of his paper.

However, there is a flaw in Lentz's proof of his shift-derivative
inequality, which is not generally true (neither is Eq. (\ref{Eq:ShiftVectorIneq})).
We now review Lentz's argument.

We begin with the shift-vector components. These are derived in Appendix
\ref{sec:B-deriveNxyz} and are similar to those in Lentz's Eqs. (19)
and (20) respectively. 
\begin{equation}
\begin{aligned}N_{x} & =\partial_{x}\phi=\mathrm{sgn}(x)\partial_{s}\phi=\frac{1}{4}\mathrm{sgn}(x)\int ds'\mathrm{sgn}(s-s')\rho(s',z-|s-s'|),\\
N_{y} & =\partial_{y}\phi=\mathrm{sgn}(y)\partial_{s}\phi=\frac{1}{4}\mathrm{sgn}(y)\int ds'\mathrm{sgn}(s-s')\rho(s',z-|s-s'|),\\
N_{z} & =\partial_{z}\phi=-\frac{1}{4}\int ds'\rho(s',z-|s-s'|),
\end{aligned}
\label{Eq:Nxyz}
\end{equation}

If we take the $z$ derivatives of both sides, and then the absolute
values, we obtain 
\begin{equation}
\begin{aligned}|\partial_{z}^{2}\phi| & =\frac{1}{4}\left|\int ds'\partial_{z}\rho(s',z-|s-s'|)\right|\\
|\partial_{z}\partial_{x}\phi| & =\frac{1}{4}\left|\int ds'\text{sgn}(s-s')\partial_{z}\rho(s',z-|s-s'|)\right|\\
|\partial_{z}\partial_{y}\phi| & =|\partial_{z}\partial_{x}\phi|
\end{aligned}
\label{Eq:|dzNzx|}
\end{equation}
When comparing the first two equations above, we note that the integrands
differ only by $\text{sgn}(s-s')$. Because $1\ge\text{sgn}(s-s')$
it might be tempting to think that the first integral is greater than
the second. However, such an inequality generally does not hold. We
demonstrate this in Appendix \ref{sec:New-SDineq} by providing a
counterexample.

\subsubsection{The energy non-negativity condition}

\begin{figure}[H]
\centering \includegraphics[scale=0.72]{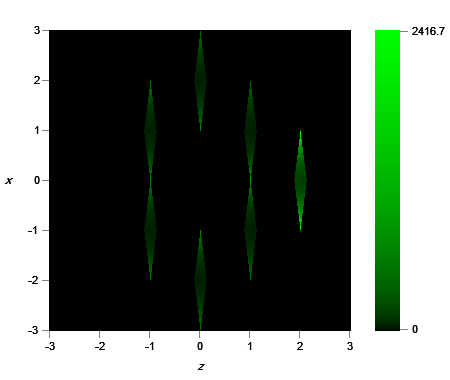} \caption{$\mathbf{\rho_{\text{L}}\partial_{z}^{2}\phi_{\text{L}}}$ for $y=10^{-6}$
($\gamma=0.43$)}
\label{Fig:Src_LXd2zNz} 
\end{figure}

A correct energy-density inequality can be stated as follows: if $\phi$
is a solution of Eq. (\ref{Eq:DE}) and if $(\partial_{z}^{2}\phi)^{2}\ge(\partial_{z}\partial_{s}\phi)^{2}$,
then 
\begin{equation}
E\ge\frac{1}{8\pi}\left(T(x,y,z)+\rho(s,z)\partial_{z}^{2}\phi\right).\label{Eq:EDIneq}
\end{equation}
We refer to this as Lentz's ``modified non-negativity inequality".
This resembles Lentz's Eq.(21) (up to an overall factor of $\frac{1}{8\pi})$,
except for the term $T(x,y,z)$ which is $0$ almost everywhere.\footref{note1}
Eq. (\ref{Eq:EDIneq}) can be derived as follows: 
\begin{itemize}
\item From Eq. (\ref{subsec:EnergyDensity}), $E=\frac{1}{8\pi}\left[T(x,y,z)+U(x,y,z)\right]$ 
\item From Eq. (\ref{Eq:Uxyz}), 
\begin{equation}
U(x,y,z)=\rho(s,z)\partial_{z}^{2}\phi+\alpha,\label{Eq:Uexpansion}
\end{equation}
where $\alpha=2\left(\left(\partial_{z}^{2}\phi\right)^{2}-\left(\partial_{z}\partial_{s}\phi\right)^{2}\right)$ 
\item Therefore $E=\frac{1}{8\pi}\left(T(x,y,z)+\rho(s,z)\partial_{z}^{2}\phi+\alpha\right)$ 
\item According to Eq. (\ref{Eq:ShiftVectorIneq}), $\alpha\geq0$, thus
proving Eq. (\ref{Eq:EDIneq}). 
\end{itemize}
The inequality Eq. (\ref{Eq:EDIneq}) yields a general non-negativity
condition: If the RHS is non-negative, then the Eulerian energy density
$E$ will be non-negative, provided that $\phi$ is a solution of
Lentz's differential equation and that the shift-derivative equality
is obeyed. We see in Fig. (\ref{Fig:Src_LXd2zNz}), for $\phi_{L}$,
the RHS is positive. However, $\phi_{L}$ is \textbf{not} a solution
of the Lentz differential equation. Among other problems, it has the
wrong sign. Therefore, we can conclude nothing from Eq. (\ref{Eq:EDIneq}),
about the sign of the energy density for the $\phi_{L}$ geometry.

Lentz's original non-negativity equation\footnote{See Eq. (21) of Lentz \cite{Lentz}}
only contains the second term, $\frac{1}{8\pi}\rho(s,z)\partial_{z}^{2}\phi$
of the modified energy inequality. However, even if the term is positive,
it could happen that the first term, $\frac{1}{8\pi}T(x,y,z)$ has
regions of negativity. We show an example of this in the next section,
where we consider a rhomboidal source.

\section{A modified version of Lentz's geometry}

\label{sec:phiMod}

\subsection{ The modified potential, $\phi_{\text{rh}}$}

\label{subsec:LentzPotential}

Consider the HP potential $\phi_{\text{rh}}$ obtained from Eq. ({\ref{Eq:phiRSVP}})
with the source function $\rho_{\text{rhomb}}(s,z(t))$ defined in
Eqs. (\ref{Eq:srcRhom}) -- (\ref{Eq:chgPflRhom}). 
\begin{equation}
\phi_{\text{rh}}(x,y,z(t))=-\frac{1}{4}v_{h}\int_{-\infty}^{+\infty}\int_{-\infty}^{+\infty}ds'dz'\Theta(z(t)-z'(t)-\frac{1}{v_{h}}|s(x,y)-s'|)\rho_{\text{rhomb}}(s',z'(t)),\label{Eq:phiMod}
\end{equation}
where $\rho_{\text{rhomb}}$ is defined by Eqs. (\ref{Eq:srcRhom})
-- (\ref{Eq:chgPflRhom}). As before, we continue with $v_{h}=1$.

$\phi_{\text{rh}}$ resembles $\phi_{L}$ in Eq. (\ref{Eq:phiLentz})
which has the same source $\rho_{\text{rhomb}}$. However, unlike
$\phi_{L}$, $\phi_{\text{rh}}$ satisfies the Lentz differential
equation, Eq. (\ref{Eq:de}), \textit{almost everywhere}.\footref{note1}.
This was discussed in section (\ref{subsec:DEsolution}). We refer
to $\phi_{\text{rh}}$ as the ``rhomboidal-source potential", and
distinguish it from $\phi_{L}$ which is also based on a rhomboidal
source but uses a different defining equation.

The rhomboidal source $\rho_{\text{rh}}(s,z)$ is shown in Fig. (\ref{Fig:Src_rhom})
if coordinate $x$ is replaced by coordinate $s$. The rhomboidal-source
shift-vector components $N_{z}^{\text{rh}}$ and $N_{x}^{\text{rh}}$,
evaluated on slice $y=0$, have the same values as $N_{z}^{L}$ and
$N_{x}^{L}$, as shown in Figs. (\ref{Fig:Nz_rhom}) and (\ref{Fig:Nx_rhom})
respectively.

In our opinion, $\phi_{\text{rh}}$ as defined in Eq. (\ref{Eq:phiMod})
is what Lentz intended in his study \cite{Lentz} instead of $\phi_{L}$
as defined in Eq. (\ref{Eq:phiLentz}). These potential function definitions
differ from one another in three ways. 
\begin{itemize}
\item In the definition of $\phi_{\text{rh}}$, we used $s$ to force symmetry
between $x$ and $y$. 
\item $\phi_{\text{rh}}$ has an overall factor $v_{h}$ whereas $\phi_{L}$
has a factor $1/v_{h}$. When $v_{h}=1$ the distinction is hidden. 
\item The overall sign is different. 
\end{itemize}
Before continuing, it is important to address the fact that, at certain
values of $(x,y,z)$, the shift vector or its derivatives may be undefined.
This is because the source function $\rho_{\text{rhomb}}$ in Eq.
(\ref{Eq:srcRhom}), has regions of discontinuity (from the Heaviside
function $\Theta$) and the source function has an argument with discontinuous
derivatives $(|x|+|y|)$. At these discontinuities, derivatives were
undefined. The shift vector is initially defined only on the domain
of points $(x,y,z)$, where the derivatives are well defined. We then
extended the domain of the shift vector and its derivatives by defining
them on the remaining points. This ``extended" (also known as ``completed")
shift vector isn't unique. What does this mean physically? The true
geometry of the universe is unlikely to have real mathematical discontinuities
of the kind that appear in the Lentz warp drive. However, for the
physical quantities computed in this study, all discontinuous behaviors
are restricted to either step or delta functions, both of which can
be treated as mathematical distributions. In particular, this implies
that our results and analysis can be applied to any smooth geometry
that matches ours outside of a proper-distance $\epsilon$ from the
discontinuities. As for the interpolation region -- i.e., the points
within $\epsilon$ of the discontinuities -- for sufficiently small
$\epsilon$, the integrated energy density will either approach $0$
for a step-function discontinuity, or will have the same sign as the
coefficient of the delta function. The planes $(x=0)$, $(y=0)$ are
physically important because they form part of the domain of the discontinuities
discussed above and ultimately lead to delta functions that contribute
to the total energy density.

\subsection{ The energy}

\label{subsec: Energy}

We now analyze the energy density of the $\phi_{\text{rh}}$ geometry
and show that it violates the WEC. Along the way, we will show, using
the $\phi_{\text{rh}}$ potential, some of the relationships discussed
in the previous section on properties of HP warp drives.

Recall Eqs (\ref{Eq: EnergyDensity} -- \ref{Eq:Uxyz}) 
\[
E=\frac{1}{8\pi}\left[T(x,y,z)+U(x,y,z)\right]
\]
where, using Eq. (\ref{Eq:Sxy}) to expand $S(x,y,z)$ 
\begin{equation}
T(x,y,z)=2\left(\delta(x)+\delta(y)\right)\partial_{s}\phi\left(\partial_{s}^{2}\phi+\partial_{z}^{2}\phi\right)+4\delta(x)\delta(y)\left(\partial_{s}\phi\right)^{2}.\label{Eq:T}
\end{equation}
and 
\[
U(x,y,z)=\left(\rho(s,z)+2\partial_{z}^{2}\phi\right)\partial_{z}^{2}\phi-2\left(\partial_{z}\partial_{s}\phi\right)^{2}.
\]

\subsubsection{ The $U$ term}

\label{subsubsec: U}

\begin{figure}[H]
\centering \includegraphics[scale=0.72]{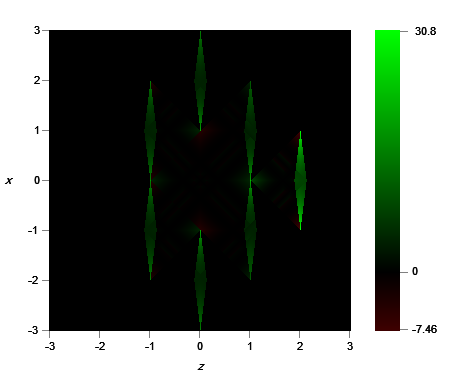} \caption{$\mathbf{|\partial_{z}^{2}\phi_{\text{rh}}|-|\partial_{z}\partial_{s}\phi_{\text{rh}}|}$
for $y=0$ (no $\gamma$-correction)}
\label{Fig:Shift-DerivIneq} 
\end{figure}

Lentz, in his Eq. (21), includes only the $U$ term (Eq. (\ref{Eq:Uxyz}))
in the energy inequality. He then concluded that the positivity of
$U$ implies the positivity of the energy density. Recall from Eq.
(\ref{Eq:Uexpansion}) that 
\[
U(x,y,z)=\rho(s,z)\partial_{z}^{2}\phi+\alpha,
\]
where $\alpha=2\left(\left(\partial_{z}^{2}\phi\right)^{2}-\left(\partial_{z}\partial_{s}\phi\right)^{2}\right)$.
$\alpha$ is positive if and only if $\phi_{\text{rh}}$ obeys the
shift-derivative inequality, Eq. (\ref{Eq:ShiftVectorIneq}). If this
would have been the case, then $U(x,y,z)\ge\rho(s,z)\partial_{z}^{2}\phi$.
This result would be equivalent to that of Lentz's Eq. (13). However,
we numerically demonstrate that the shift-derivative inequality fails.
We portray this by plotting the term $\mathbf{|\partial_{z}^{2}\phi_{\text{rh}}|-|\partial_{z}\partial_{s}\phi_{\text{rh}}|}$.
It has regions of negativity, so the shift-derivative inequality fails.
The results are shown in Fig. (\ref{Fig:Shift-DerivIneq}). 

This image appears to show that the $\phi_{\text{rh}}$ obeys the
shift-derivative inequality. However, we can see a hint of the small
negative contributions. Therefore we replotted the function, but with
$\gamma$ set to $0.43$, so that small negative values were sufficiently
amplified to be visible. 
\begin{figure}[H]
\centering \includegraphics[scale=0.72]{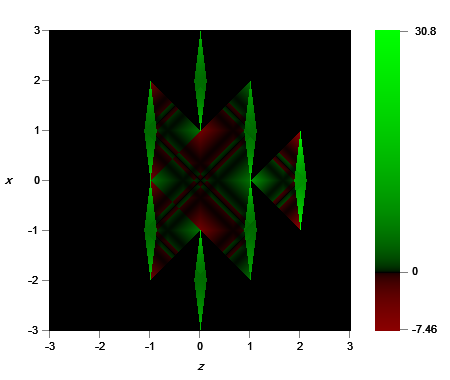}
\caption{$\mathbf{|\partial_{z}^{2}\phi_{\text{rh}}|-|\partial_{z}\partial_{s}\phi_{\text{rh}}|}$
for $y=0$ ($\gamma=0.43$)}
\label{Fig:Shift-DerivIneq_gma=00003D00003D00003D00003D00003D00003D0.43} 
\end{figure}

This new image clearly demonstrates the failure of the shift-derivative
inequality for our rhomboidal-source geometry.

\begin{figure}[H]
\centering \includegraphics[scale=0.72]{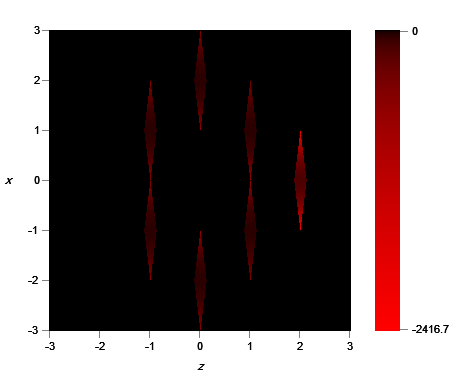} \caption{$\mathbf{\rho_{\text{rh}}\partial_{z}^{2}\phi_{\text{rh}}}$ for $y=10^{-6}$
($\gamma=0.43$)}
\label{Fig:Src_rhXd2zNz} 
\end{figure}

We have yet to consider the first term, $\rho(s,z)\partial_{z}^{2}\phi$,
on the RHS of Eq. (\ref{Eq:Uexpansion}). Even though the second term
has regions of negativity, their values are small, so can, in principle,
be easily compensated by large positive values of the first term.
However, the term $\rho(s,z)\partial_{z}^{2}\phi$ is displayed in
Fig. \ref{Fig:Src_rhXd2zNz} and can immediately be seen to have values
$\le0$.

Note that this function has the opposite sign to that shown in Fig.
\ref{Fig:Src_LXd2zNz} for Lentz's original potential. These figures
provide critical insights into the failure of Lentz's demonstration
of a WEC-compliant warp drive. Even if one was to set aside concerns
about sets\footref{note1} of measure 0 and the shift-derivative
inequality, the key premise of Lentz's demonstration was the non-negativity
of $\rho_{\text{L}}\partial_{z}^{2}\phi_{\text{L}}$. Owing to a sign
mistake in the solution of his differential equation, Lentz incorrectly
concluded that $\mathbf{\rho_{\text{L}}\partial_{z}^{2}\phi_{\text{L}}}\ge0$.

It is possible that the modified potential $\rho_{\text{rh}}$ (which
solves Lentz's differential equation almost everywhere) could have
led to the same positivity as $\rho_{\text{L}}$, but as we can see
from the figure, $\rho_{\text{rh}}\partial_{z}^{2}\phi_{\text{rh}}$
is everywhere non-positive.

\subsubsection{The $T$ term}

\label{subsec:T}

\begin{figure}[H]
\centering \includegraphics[scale=0.72]{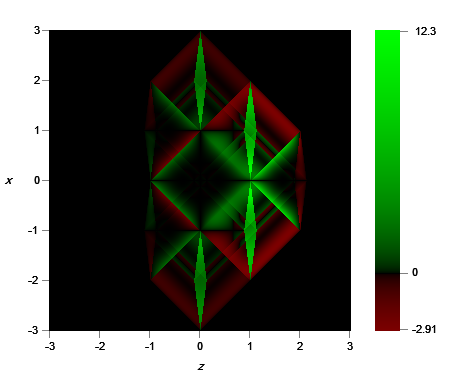} \caption{$\partial_{s}\phi_{\text{rh}}\left(\partial_{s}^{2}\phi_{\text{rh}}+\partial_{z}^{2}\phi_{\text{rh}}\right)$
in the plane $x=0$ ($\gamma=0.5$) }
\label{Fig:T-termFactor} 
\end{figure}

The energy density inequality in Eq. (\ref{Eq: EnergyDensity}) includes
the term $T$, given in Eq. (\ref{Eq:T}). Formally, this term is
irrelevant except for the singular planes $x=0$ and $y=0$. However,
as previously mentioned, $T$ should be understood as a distribution.
This can be regarded as a (non-unique) approximation, where the $\delta$
functions are replaced by positive normalized spikes. In this case,
note that the second term of $T(x,y,z)$ -- namely, $4\delta(x)\delta(y)\left(\partial_{s}\phi\right)^{2}$
-- is non-negative, albeit only near the singular line. On the other
hand, the first term $2(\delta(x)+\delta(y))\partial_{s}\phi\left(\partial_{s}^{2}\phi+\partial_{z}^{2}\phi\right)$
need not be non-negative. To illustrate this, we show in Fig. (\ref{Fig:T-termFactor})
the partial negativity of this term for our rhomboidal geometry.

In summary, $T(x,y,z)$ might be partially negative and thus, near
the singular lines, the Eulerian energy density could become negative
regardless of the values of $U$. We cannot ignore the singular regions
just because of the delta functions. The WEC does not have a `singular-line
exclusion.'

In fact, as we shall discuss further in the context of ``no-go theorems,"
the $T$ term is essential for computing the total energy that is
the integrated energy-density. Even though the energy density is ambiguous
(owing to the options for completion of the singular\textcolor{green}{{}
}planes), the integrated value (with unambiguous treatment of $\delta$
functions) is not. The total energy has a non-zero contribution from
both $T$ and $U$.

However, if the RHS has some regions of negativity -- as it does
for $\phi_{\text{rh}}$ (see Fig. \ref{Fig:Src_rhXd2zNz})-- then
nothing can be concluded about the non-negativity of $E$.

\subsubsection{ Energy density}

\label{subsubsec: E_density}

In summary, we have shown that both terms comprising the energy density
term $U$ have regions of negativity, and by comparing their respective
figures, we can easily see that their total (i.e., $U(x,y,z)$) also
has regions of non-negativity. In particular, these regions are outside
the critical planes $x=0$ and $y=0$ so cannot be compensated by
the $T$ term that we considered in Section \ref{subsec:T}.

\begin{figure}[H]
\centering \includegraphics[scale=0.72]{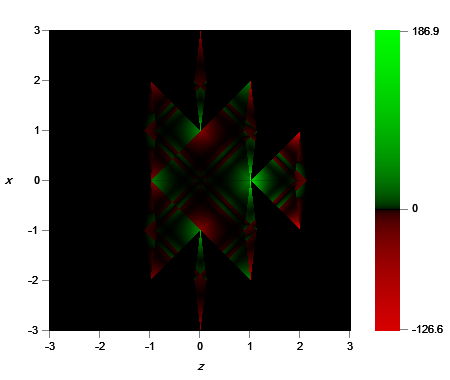} \caption{The Eulerian energy density $E_{\text{rh}}$ for the rhomboidal source
function at $y=10^{-6}$ ($\gamma=0.43$)}
\label{Fig:E_rh} 
\end{figure}

In Fig. (\ref{Fig:E_rh}), we show the Eulerian energy density $E_{\text{rh}}$
away from the critical planes, that is, we display $U(x,y,z)$ using
Eq. \ref{Eq:Uxyz}. This appears to differ significantly from the
Eulerian energy density based on $\phi_{L}$, as shown in Fig. (\ref{Fig:E_L_rhom}).
Nevertheless, $E_{\text{rh}}$ is also negative in some regions and
thus, the $\phi_{\text{rh}}$ geometry also violates the Weak Energy
Condition.

There is a significant difference between $E_{L}$ and $E_{\text{rh}}$\textcolor{green}{{}
}related to the discussion on energy density in Section \ref{subsec:phiLentz}.
Computing $E_{L}$ via (\ref{Eq:extrinsic}) and (\ref{Eq:Eulerian1})
- the correct energy density corresponding to $\phi_{L}$ - reveals
$45^{\circ}$ bands that extend to infinity.\textcolor{green}{{}
}Thus, the total energy of the $\phi_{L}$ warp-spacetime is infinite
and therefore violates a key requirement of the geometry. This is
a consequence of the fact that Lentz \cite{Lentz} defined $\phi_{L}$
as a function of a source that has incorrect arguments (see Appendix
\ref{subsec:A.1}) and, as a consequence, is not symmetric with respect
to $x$ and $y$\textcolor{green}{.} However, this is not the case
for the $\phi_{\text{rh}}$ warp-spacetime, for which the energy is
bounded. This can be proven from the fact that the rhomboidal source
(with the correct arguments) satisfies the three conditions of Subsection
\ref{subsec:Finiteness}. The last two conditions (source-domain boundedness
and $s(x,y)=|x|+|y|$) are satisfied trivially. To demonstrate the
cancellation condition, consider the seven rhomboidal (sub)domains
of the source function. Any diagonal line either intersects no rhomboids
or intersects rhomboids whose weights add up to $0$ and whose intersection
line-segments are identical to one another relative to their respective
rhomboid centroids.

There is also a significant difference between the Eulerian energy
densities in Fig. 3 in ref.\textcolor{green}{{} }\cite{Lentz} and
$E_{\text{rh}}$. This is because computing\textcolor{green}{{} }$E_{L}$
using Lentz's Eq. (17) yields a completely different result than that
obtained using Eqs. (\ref{Eq:extrinsic}) and (\ref{Eq:Eulerian1}),
respectively. Note that computing $E_{\text{rh}}$ using Eq. (\ref{Eq: EnergyDensity})
with $U(x,y,z)$ given by Eq. (\ref{Eq:Uxyz}), which is equivalent
to Lentz's Eq. (17) away from the critical planes, yields the same
result as obtained using Eqs. (\ref{Eq:extrinsic}) and (\ref{Eq:Eulerian1}),
respectively. This distinction is made because $\phi_{\mathrm{rh}}$
is a solution of Lentz's hyperbolic differential equation, whereas
$\phi_{L}$ is not.

\section{Analysis of a rectangular-source HP warp drive}

\label{sec:SrcRect}

Next, we next consider a slight modification of the Lentz source function.
Instead of a collection of rhomboids, we take the source to consist
of a collection of rectangles. \footnote{Although function $g_{i}(s)$ presented in Eq. (\ref{Eq:srcRect})
may seem trivial, it is provided so that equation fits the general
scheme presented in Appendix \ref{sec:C-evalIntegrals}.}

\begin{equation}
\rho_{\text{rect}}(s,z)=\tilde{\rho}_{\text{rect}}(x,y,z)=W\sum_{i=1}^{4}\tilde{\alpha}_{i}\Theta\left(p_{1,i}(z)\right)\Theta\left(p_{2,i}(s)\right)g_{i}(s),\label{Eq:srcRect}
\end{equation}
where 
\begin{equation}
\begin{aligned}p_{1,i}(z) & =\tilde{L}_{z}-|z-\xi_{i}|,\\
p_{2,i}(s) & =\tilde{L}_{x}-||s|-\beta_{i}|,\\
g_{i}(s) & =Q_{\max}.
\end{aligned}
\label{Eq:shapeRect}
\end{equation}
where after substituting rectangles for rhomboids, the parameter descriptions
and values are the same as those described for the rhomboidal sources
in Section \ref{subsec:SrcRhom}, except that the minimum charge $Q_{\min}$
is not relevant. The constraints on the overall source configuration
are also the same.

\begin{figure}[H]
\centering \includegraphics[scale=0.72]{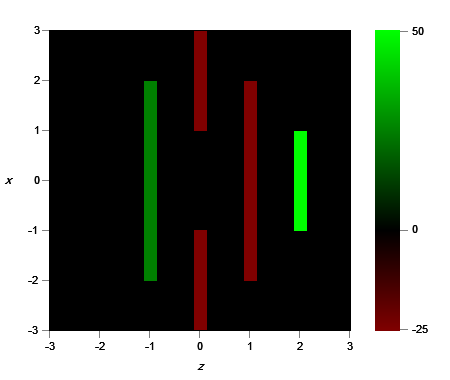} \caption{Unscaled rectangular source $\rho_{\text{rect}}N_{z}^{\text{rect}}(0,0,0)$}
\label{Fig:Src_rect} 
\end{figure}

The unscaled rectangular source is displayed in Fig.(\ref{Fig:Src_rect}).
That is, we show $\rho_{\text{rect}}N_{z}^{\text{rect}}(0,0,0)$,
for an easy comparison with the unscaled rhomboidal source shown in
Fig. \ref{Fig:Src_rhom}. 

Similarly to Eq. (\ref{Eq:phiMod}), we obtain the shift-vector potential
$\phi_{\text{rect}}$. 
\begin{equation}
\begin{aligned}\phi_{\text{rect}}(x,y,z(t)) & =-\frac{1}{4}\int_{-\infty}^{\infty}dx'dz'\Theta\left(z(t)-z'-|x-x'|\right)\rho_{\text{rect}}(s',z'),\end{aligned}
\label{Eq:phiRect}
\end{equation}
where $z(t)=z-vt$, $v$ is the speed of the warp drive, and $\rho_{\text{rect}}$
is defined by Eqs. (\ref{Eq:srcRect}) -- (\ref{Eq:shapeRect}).
Recall that $s(x,y)=|x|+|y|$ and $\rho(s(x,y),z)=\rho(-s(x,y),z)=\tilde{\rho}(x,y,z)$.
As in the rhomboidal case, we have a slight abuse of notation, and
set $\phi_{\text{rect}}(s,z)=\phi_{\text{rect}}(x,y,z)$.

\begin{figure}[H]
\centering \includegraphics[scale=0.65]{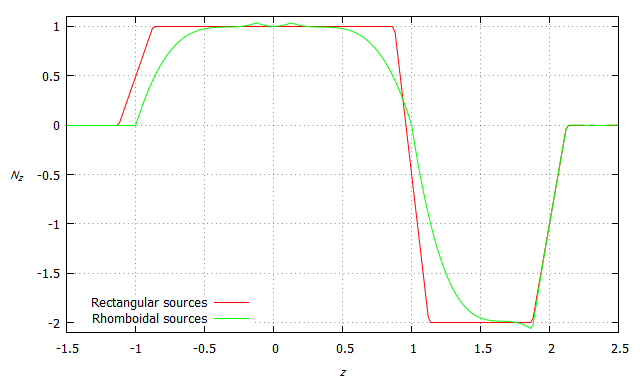}\caption{Comparative profiles\textcolor{green}{{} }of $N_{z}^{\text{rect}}(x,y,z)$
and $N_{z}^{\text{rhom}}(x,y,z)$ at fixed $(x,y)=(0,0).$}
\label{Fig:NzPfl_rhom_rect} 
\end{figure}

\begin{figure}[H]
\centering \includegraphics[scale=0.65]{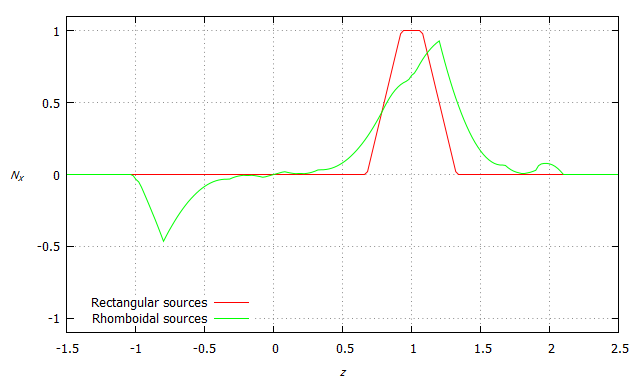}\caption{Comparative profiles of $N_{x}^{\text{rect}}(x,y,z)$ and $N_{x}^{\text{rhom}}(x,y,z)$
at fixed $(x,y)=(-0.2,0).$}
\label{Fig:NxPfl_rhom_rect} 
\end{figure}

As mentioned in Section \ref{subsec:SrcRhom}, the warp drive based
on rectangular sources is superior to that based on rhomboidal sources,
in the sense that $N_{z}^{\text{rect}}$ is non-zero and precisely
level. and $N_{x}^{\text{rect}}$ and $N_{y}^{\text{rect}}$ are precisely
zero over a broader central region. This is illustrated in Fig. \ref{Fig:NzPfl_rhom_rect},
which shows the comparative profiles of $N_{z}^{\text{rect}}$ and
$N_{z}^{\text{rhom}}$, along the $z$-direction at $(x,y)=(0,0)$,
and in Fig. \ref{Fig:NxPfl_rhom_rect}, which shows the comparative
profiles of $N_{x}^{\text{rect}}$ and $N_{x}^{\text{rhom}}$, also
along the $z$-direction, but at $(x,y)=(-0.2,0)$.

\begin{figure}[H]
\centering \includegraphics[scale=0.72]{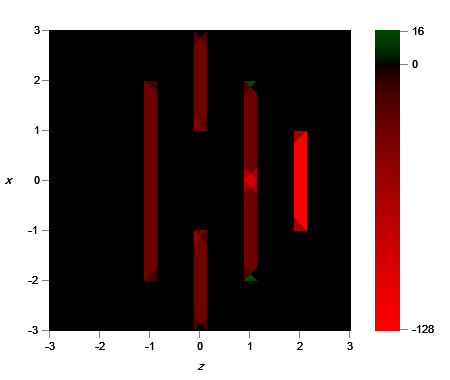} \caption{$\mathbf{\rho_{\text{rect}}\partial_{z}^{2}\phi_{\text{rect}}}$($\gamma=0.60$)}
\label{Fig:Src_rectXd2zNz} 
\end{figure}

Does the resultant warp-drive geometry have a positive WEC? We will
briefly see by direct computation that this does not occur. First,
let us examine what we would expect based on Lentz's inequality analysis.
Recall the argument in section \ref{subsec:EulerLentz}, that if $T_{\text{rect}}(x,y)+\rho_{\text{rect}}\partial_{z}^{2}\phi_{\text{rect}}$
were positive (where $T_{\text{rect}}(x,y)$ is zero almost everywhere),
then the energy density would be positive. We observed that this conclusion
hinges on the validity of the shift-derivative inequality. However,
that inequality turns out to be irrelevant (just as in the rhomboidal
case) because, as shown in Fig. (\ref{Fig:Src_rectXd2zNz}), $\rho_{\text{rect}}\partial_{z}^{2}\phi_{\text{rect}}$
is negative in the rectangular regions where the source is non-zero.

\begin{figure}[H]
\centering \includegraphics[scale=0.72]{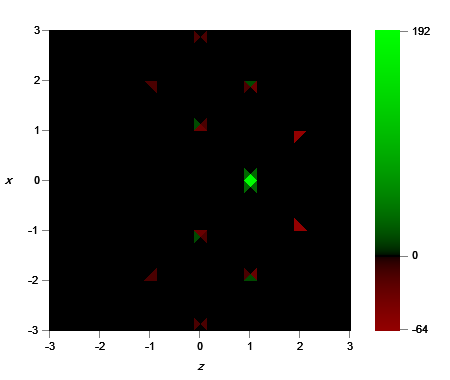} \caption{The Eulerian energy density $E_{\text{rect}}$ for the rectangular
source function ($\gamma=0.50$)}
\label{Fig:E_rect} 
\end{figure}

What we have demonstrated thus far, is that the rectangular-source
geometry doesn't meet the requirements of Lentz's modified energy
inequality theorem for proving non-negativity of the Eulerian energy
density, not even approximately. Of course, it could still happen
``accidentally'' that the Eulerian energy density turns out to be
non-negative. In Fig. (\ref{Fig:E_rect}) we show the Eulerian energy
density $E_{\text{rect}}$ computed using Eq. (\ref{Eq:Uxyz}). As
for the rhomboidal-source example, it is negative in some regions
and thus violates the Weak Energy Condition. It is noteworthy that,
unlike the rhomboidal-source example, the energy-density is zero except
in a few small regions. However, we don't see how to modify the sources
to eliminate regions of negative energy-density altogether. Also,
even if we could find such sources, we would need to separately examine
the contribution of $T$ which could turn out to be negative.

\section{Applicability of no-go theorems to Lentz-style warp drives}

\label{sec:no-go}

In this study, we address an apparent paradox in the literature. On
the one hand, there are ``no-go" theorems stating that Natario warp
drives must violate various energy conditions, such as the WEC. On
the other hand, there is an example, proposed by Lentz, of a Natario
warp drive that supposedly has been demonstrated to satisfy the WEC.
There are only three possible ways to resolve this paradox: 
\begin{enumerate}
\item The no-go theorems are wrong. 
\item The Lentz demonstration is wrong. 
\item Lentz's warp drive does not satisfy the conditions required for the
no-go proofs. 
\end{enumerate}
Unfortunately, lacking a resolution to this issue, there has continued
to be murkiness about whether a WEC-compliant Natario warp drive can
be constructed.

It was tempting to assume no errors either in the no-go proofs or
in Lentz's demonstration, and to focus instead on whether Lentz's
warp drive satisfies the assumptions of the no-go theorems. However,
it has turned out that this level of subtlety is unnecessary. Lentz's
demonstration is incorrect. Hence, the paradox is resolved: no Natario
warp-drive has been demonstrated to comply with the WEC. Hopefully,
this conclusion will help dispel the murkiness of the status of the
Lentz drive.

There is still the question of whether one can create a Natario warp
drive by relaxing any of the conditions of the no-go theorems. In
particular, we considered the theorems of Santiago et al. \cite{Santiago}.
In Section 7.4, the authors proved that a Natario warp-drive must
violate the WEC. The authors derived, for a 0-vorticity warp drive
(such as the Lentz warp drive) (Eq. 7.12 of ref. \cite{Santiago}),
the Eulerian energy density $E$ satisfies $E=\partial_{i}\mathcal{S}_{i}=0$
where $\mathcal{S}_{i}=\frac{1}{16\pi}\left(N_{i}\partial_{j}N_{j}-N_{j}\partial_{i}N_{j}\right)$.
From Stoke's theorem, $\int d^{3}xE=0$, provided that $\mathcal{S}_{i}$
falls off quickly enough so that integral of $\mathbf{\mathcal{S}}_{i}$
is $0$ on the surface at infinity. From this, we see that the energy
density is either precisely zero everywhere or must have some regions
of negativity (of equal weight to regions of positivity), hence violating
the WEC.

Note that if $\mathcal{S}$ were to fall off quadratically, then the
boundary-integral would be non-zero and the total energy would be
finite. Finiteness is important because we would still have a physically
meaningful theory, but the above WEC-violation proof would fail. Unfortunately,
WEC compliance cannot be established simply by showing non-negativity
of the Eulerian energy density. One must show the non-negativity of
the energy density for \textbf{any} reference frame. One way to do
this, as argued in Ref. \cite{Santiago}, is to show that in addition
to the non-negativity of the Eulerian energy density, there is also
non-negativity of a second parameter which is the sum of the average
Eulerian pressure and Eulerian energy density. The authors prove in
their section 7.4 (in the discussion of the Null Energy Condition)
that this second parameter must have some regions of negativity. This
proof does not appear to depend on the rapid spatial fall-off of $\mathcal{S}_{i}$.
However, the proof requires an additional (physically motivated) assumption:
the warp drive must be turned on at large negative times and turned
off at large positive times. We refer to this as the ``on-off condition."

In summary, Santiago et al. \cite{Santiago} proved that the WEC is
violated provided that $\mathcal{S}_{i}$ falls off quickly enough
or that the warp drive satisfies the on-off condition.\footnote{ Santiago et al. \cite{Santiago} prove that the on-off condition
implies NEC violation, which implies WEC violation.} If we are willing to relax these conditions, then it might be possible
to construct a WEC-compliant Natario warp drive. It should be emphasized
that we did not find such a warp drive (although we have investigated
a Fell-Heisenberg drive \cite{Fell} with a positive Eulerian density,
which nevertheless\textcolor{green}{{} }violates the WEC), nor have
we proven that such a WEC-compliant drive can be constructed.

\subsection{ Validity of Stoke's theorem for $\phi_{\text{rh}}$ and $\phi_{\text{rect}}$}

\label{Stokesvalidity}

There is one more set of assumptions implicit in the above theorems,
which are \uline{not} satisfied by Lentz's warp drive or the variants
explored in this paper. These assumptions have to do with the smoothness
of the geometric parameters, a quality we expect from physical theories.
The proposals we explored all had regions where the shift vector was
not differentiable. This lack of differentiability could, in principle,
render the theorems (for example, Stoke's theorem) invalid and could
therefore provide a loophole that would allow WEC-compliance. One
reason we have gone through the computation and analysis of the Lentz
warp drives, is to determine whether there may be a possible role
of non-differentiability in the non-applicability of the no-go theorems.

The Stokes'-based no-go proof is probably valid despite the lack of
differentiability of the $\phi_{\text{rh}}$ and $\phi_{\text{rect}}$
geometries. To observe this, we computed the total energy $E_{\text{TOT}}$
for both geometries, where from Eq. (\ref{Eq: EnergyDensity}), 
\begin{equation}
E_{\text{TOT}}=\frac{1}{8\pi}\int dxdydz\left(T(x,y,z)+U(x,y,z)\right).\label{Eq: TotalEnergy}
\end{equation}
See Appendix \ref{sec:EstEtot} for details on how to integrate the
$T$ term.

\begin{figure}[H]
\centering \includegraphics[scale=0.5]{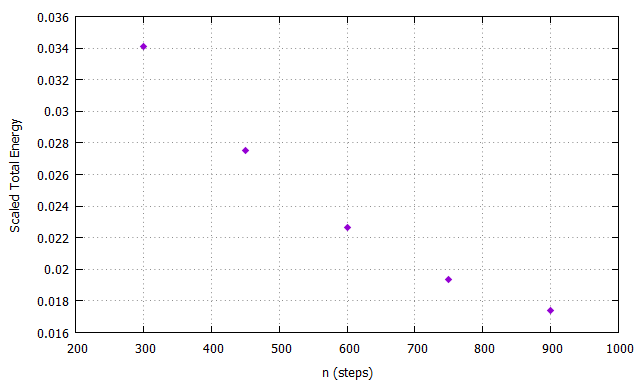}
\caption{$\frac{E_{\text{TOT}}^{\text{rh}}}{E_{\text{scale}}^{\text{rh}}}$
as a function of $n$ for the rhomboidal source function.}
\label{Fig: Etotrhomb} 
\end{figure}

\begin{figure}[H]
\centering \includegraphics[scale=0.5]{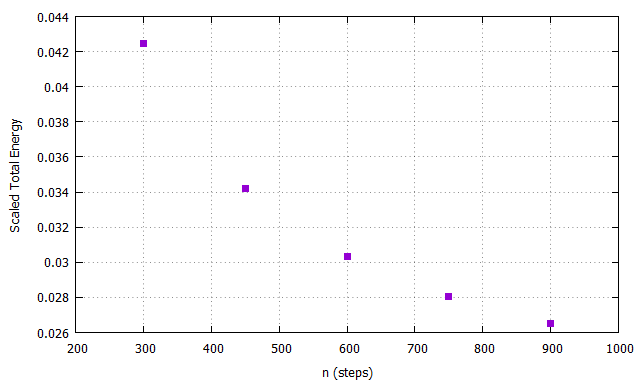}
\caption{$\frac{E_{\text{TOT}}^{\text{rect}}}{E_{\text{scale}}^{\text{rect}}}$
as a function of $n$ for the rectangular source function.}
\label{Fig:Etotrect} 
\end{figure}

Because our goal is to compare the total energy to $0$, we need to
scale the energy-density by a constant $1/E_{\text{scale}}$, which
we define as 
\begin{equation}
E_{\text{scale}}=\int dxdydz\left(|T(x,y,z)|+|U(x,y,z)|\right).
\end{equation}
(Note that because $T$ involves the absolute values of $\delta$-functions,
we adopt the definition that $|\delta(x)|\equiv\delta(x)$. ) As an
example, we numerically computed using a grid with 300 steps in each
direction, $E_{\text{TOT}}^{\text{rh}}/E_{\text{scale}}^{\text{rh}}\approx0.034$.
This calculation was numerically consistent with a total energy of
$0$. For better numerical accuracy, we must increase $n$ -- the
number of grid steps in each direction. In Figs. \ref{Fig: Etotrhomb}
and \ref{Fig:Etotrect}, we show the scaled total rhomboidal geometry
and rectangular geometry energies as a function of $n$. The graphs
are consistent with the convergence to $E_{\text{TOT}}=0$.

\section{Conclusions}

\label{sec:Conclusions}

In ref. \cite{Lentz}, Lentz proposed a new warp-drive geometry claimed
to be WEC compliant. In this study, we showed that this claim is incorrect.
The Lentz warp drive violates the WEC. We have shown this directly
by the numerical computation of the energy density, as shown in Fig.
\ref{Fig:E_L_rhom}. We also found and explained various errors made
by Lentz in his computations, as well as in the development of his
theoretical requirements for the construction of a WEC-compliant warp
drive.

None of this should be surprising, given that there are no-go theorems
discussed in Section \ref{sec:no-go}, which state the impossibility
of WEC-compliant warp drives. Of course, similar to all theorems,
there are assumptions and conditions. These include the differentiability
requirements violated by Lentz's geometry.

We also explored whether Lentz's proposal and implementation could
be modified to achieve the desired result. We were able to test a
couple of geometric models that were similar to Lentz's but that corrected
some of the original errors. Unfortunately, these violate the WEC
as well.

In the last section, we discuss the applicability of various no-go
theorems to warp drives presented in this study. Our modified rhomboidal
and rectangular warp drives satisfy the rapid-falloff condition of
the no-go theorem for Eulerian energy, but not the on-off condition
or the differentiability condition. However, despite these possible
no-go loopholes, the Lentz and modified Lentz warp drives violate
the WEC, as we have shown.

\appendix

\section{Derivation of the Hyperbolic Potential \label{sec:A-MotivPhi}}

In Section \ref{sec:phiMod}, we introduce a modified version of Lentz's
potential given by Eq. (\ref{Eq:phiMod}), and we then generalize
this for general source functions $\rho$, to Eq. (\ref{Eq:phiRSVP}).
This potential is referred to as the hyperbolic potential (HP). We
then re-introduce $v_{h}$ which had earlier been set as $v_{h}=1$,
and obtain 
\begin{equation}
\phi(x,y,z)=-\frac{1}{4}v_{h}\int_{-\infty}^{+\infty}\int_{-\infty}^{+\infty}ds'dz'\Theta(z-z'-\frac{1}{v_{h}}|s-s'|)\rho(s',z'),\label{eq:A-01}
\end{equation}
where it is implicit that $z$ is a function of $t$ and $s=|x|+|y|$.
We demonstrate via Eq. (\ref{Eq:Sxy}), that when $v_{h}=1$, $\phi$
is a solution of Eq. (\ref{Eq:de}) everywhere except for the planes
$x=0$ and $y=0$. This is also the case for general values of $v_{h}$.\textcolor{green}{{}
}Next, we explain how we obtain Eq. (\ref{eq:A-01}) as a solution
to Lentz's hyperbolic differential equation Eq.(\ref{Eq:de}) , which
is repeated here for convenience.

\begin{equation}
\begin{aligned}\partial_{x}^{2}\phi+\partial_{y}^{2}\phi-\frac{2}{v_{h}^{2}}\partial_{z}^{2}\phi & =\rho,\end{aligned}
\label{eq:A-02}
\end{equation}
(where the tildas are removed). If we now require that

\begin{equation}
\phi(x,y,z)=\phi\left(s(x,y),z\right),\label{eq:A-03}
\end{equation}
it can be shown that, by applying the chain rule to (\ref{eq:A-03})
and performing some algebra, that, if $x,y\neq0$ and $s=|x|+|y|$,
then

\begin{equation}
\partial_{x}^{2}\phi=\partial_{y}^{2}\phi=\partial_{s}^{2}\phi,\label{eq:A-04}
\end{equation}
where, from this point forward, we exclude $x=0$ and $y=0$ from
the domain of $\phi$. We may then rewrite Eq. (\ref{eq:A-02}) as

\begin{equation}
2\partial_{s}^{2}\phi-\frac{2}{v_{h}^{2}}\partial_{z}^{2}\phi=\rho.\label{eq:A-05}
\end{equation}
Then, by multiplying both sides of Eq. (\ref{eq:A-05}) by $-2/v_{h}^{2}$
and rearranging, we obtain

\begin{equation}
\partial_{z}^{2}\phi-v_{h}^{2}\partial_{s}^{2}\phi=-\frac{v_{h}^{2}}{2}\rho.\label{eq:A-06}
\end{equation}
The preceding equation is readily recognized as a 1-dimensional wave
equation, written canonically as 
\begin{equation}
\partial_{t}^{2}\phi-v^{2}\partial_{s}^{2}\phi=f\label{eq:A-07}
\end{equation}
with the identification

\begin{equation}
\begin{array}{ccc}
t & = & z\\
v^{2} & = & v_{h}^{2}\\
f & = & -\frac{v_{h}^{2}}{2}\rho.
\end{array}\label{eq:A-08}
\end{equation}
$\phi$ can therefore be written as \textcolor{green}{{} } 
\begin{equation}
\phi(t,s)=\int_{-\infty}^{+\infty}\int_{-\infty}^{+\infty}dt'ds'G(t-t',s-s')f(t',s')
\end{equation}
where $G$ is the Green's function given by

\begin{equation}
G(t,x)=\frac{1}{2v}\Theta\left(t-\frac{|s|}{v}\right)\label{eq:A-09}
\end{equation}
Then the corresponding solution to Eq. (\ref{eq:A-07}) is given by

\begin{equation}
\phi=\int_{-\infty}^{+\infty}\int_{-\infty}^{+\infty}dt'ds'\frac{1}{2v}\Theta\left(t-t'-\frac{|s-s'|}{v}\right)f\left(t',s'\right).\label{eq:A-10}
\end{equation}
Then, substituting Eqs. (\ref{eq:A-08}) into Eq. (\ref{eq:A-10})
yields

\begin{equation}
\phi=\int_{-\infty}^{+\infty}\int_{-\infty}^{+\infty}dz'ds'\frac{1}{2v_{h}}\Theta\left(z-z'-\frac{1}{v_{h}}|s-s'|\right)\left(-\frac{v_{h}^{2}}{2}\rho(s',z')\right)\label{eq:A-11}
\end{equation}
which simplifies to

\begin{equation}
\phi=-\frac{1}{4}v_{h}\int_{-\infty}^{+\infty}\int_{-\infty}^{+\infty}dz'ds'\Theta\left(z-z'-\frac{1}{v_{h}}|s-s'|\right)\rho(s',z')\label{eq:A-12}
\end{equation}
and we can observe that the RHS of Eq. (\ref{eq:A-12}) is identical
to that of Eq. (\ref{eq:A-01}).

\subsection{Why Lentz's original potential is not a hyperbolic potential}\label{subsec:A.1}

Eq. \ref{eq:A-12} can be compared with Lentz's original potential,
given by eq. (\ref{Eq:phiLentz}) and is repeated for convenience.

\begin{equation}
\begin{aligned}\phi_{L} & =\frac{1}{4v_{h}}\int_{-\infty}^{\infty}dx'dz'\Theta\left(z-z'-\frac{1}{v_{h}}|x-x'|\right)\rho(|x'|+|y|,z')\end{aligned}
\label{eq:A-13}
\end{equation}
Eq. (\ref{eq:A-13}) exhibits the following differences, relative
to Eq. (\ref{eq:A-12}): 
\begin{itemize}
\item An overall sign that is positive instead of negative. 
\item An overall factor is $1/v_{h}$ instead of $v_{h}$. 
\item A factor $|x-x'|$ instead of $|s-s'|$, in the argument to $\Theta$. 
\item The expression $|x'|+|y|$ instead of $s'$, as in the second argument
to $\rho$. 
\end{itemize}
The incorrectness in Eq. (\ref{eq:A-13}) can be obtained numerically.
Recall that we have two methods for computing the Eulerian energy
density: 1) via Lentz's Eq. (17), which requires $\phi$ tp be a solution
of Eq. (\ref{eq:A-02}); and 2) via Eqs. (\ref{Eq:extrinsic}) and
(\ref{Eq:Eulerian1}), which is correct regardless. The images of
the Eulerian energy density derived from (\ref{eq:A-12}) via method
1 are identical (other than possibly on the lines $x=0$ or $y=0$)
to those obtained from Eq. (\ref{eq:A-12}) using method 2. The same
cannot be said for Eq. (\ref{eq:A-13}). In particular, the sign error
(first bullet, above) directly leads to the incorrect image, as shown
in Fig. 3, in Ref.\textcolor{green}{{} }\cite{Lentz}, which demonstrates
that the WEC is satisfied. That is, if we begin with Eq. (\ref{eq:A-12})
and only change sign, the image (without $\gamma$-correction) of
Eulerian energy density near $y=0$, as obtained via method 1, is
qualitatively the same erroneous result, that is, different from the
correct result obtained via method 2, as shown in Lentz's\textcolor{green}{{}
}\cite{Lentz}\textcolor{green}{{} }Fig. (3).

The last two bullets imply that Eq. (\ref{eq:A-13}) is inconsistent
with Lentz's prescription that $s=|x|+|y|$. These errors lead to
violations of the boundedness requirement for the shift vector components
and Eulerian energy density, and thus to infinite energy. This is
not the case for the potential in Eq. (\ref{eq:A-12}). Furthermore,
the asymmetry between $x$ and $y$ means that $\partial_{x}^{2}\phi\neq\partial_{y}^{2}\phi$.
(See Appendix \ref{sec:F-superlumin} for a discussion of $\partial_{y}^{2}\phi$.)
The equality of these two 2nd-order partial derivatives is a vital
step, i.e. Eq. (\ref{eq:A-04}), to derive the potential: because
the equality does not hold in this case, the original differential
equation (\ref{eq:A-02}) does not reduce to Eq. (\ref{eq:A-06})
and thus cannot be identified with the 1-dimensional wave equation
(\ref{eq:A-07}); therefore, Eq. (\ref{eq:A-13}) cannot be a solution
to Eq. (\ref{eq:A-02}).

\section{Derivation of the expressions for the shift vector components}\label{sec:B-deriveNxyz}

The shift vector components are presented in Eq. (\ref{Eq:Nxyz})
and repeated below for convenience, except that, for greater generality,
we do not assume that $v_{h}=1$:

\begin{equation}
N_{z}=\partial_{z}\phi=-\frac{1}{4}v_{h}\int_{-\infty}^{\infty}ds'\rho(s',z-v_{h}^{-1}|s-s'|)\label{eq:B-01}
\end{equation}

\begin{equation}
N_{x}=\partial_{x}\phi=\mathrm{sgn}(x)\partial_{s}\phi=\frac{1}{4}\mathrm{sgn(x)}\int_{-\infty}^{\infty}ds'\mathrm{sgn}(s-s')\rho(s',z-v_{h}^{-1}|s-s'|)\label{eq:B-02}
\end{equation}

\begin{equation}
N_{y}=\partial_{y}\phi=\mathrm{sgn}(y)\partial_{s}\phi=\frac{1}{4}\mathrm{sgn(y)}\int ds'\mathrm{sgn}(s-s')\rho(s',z-v_{h}^{-1}|s-s'|)\label{eq:B-03}
\end{equation}
The first two equations are derived in Appendix \ref{subsec:B.1}
and the integral that appears in the last of these equations is derived
in Appendix \ref{subsec:B.2}.

\subsection{Derivation of $N_{z}$}

\label{subsec:B.1}

In deriving $N_{z}$, we will make use of the following identities:

\begin{equation}
\partial_{u}\Theta\left(q(u,w,...)\right)\equiv\delta\left(q(u,w,...)\right)\partial_{u}q\label{eq:B-04}
\end{equation}
and

\begin{equation}
\int_{-\infty}^{\infty}\delta\left(q(u,w,...)\right)f(u,w,...)du\equiv\sum_{i}\frac{f(r_{i},w,...)}{\left|\partial_{u}q|_{u=r_{i}}\right|}\label{eq:B-05}
\end{equation}
where $r_{i}$ is the ith root of $q(u,w,...)=0$, treating $q$ as
a function of $u$, only. Analogous equations hold for variables $w$,
etc.

As $N_{z}=\partial_{z}\phi$, where $\phi$ is given by Eq. (\ref{eq:A-12}),
we can differentiate Eq. (\ref{eq:A-12}) with respect to $z$,

\begin{equation}
N_{z}=-\frac{1}{4}v_{h}\int_{-\infty}^{\infty}\left[\int_{-\infty}^{\infty}\partial_{z}\Theta\left(z-z'-v_{h}^{-1}|s-s'|\right)\cdot\rho(s',z')dz'\right]ds'.\label{eq:B-06}
\end{equation}
Thus, we must evaluate the derivative $\partial_{z}\Theta\left(z-z'-v_{h}^{-1}|x-x'|)\right)$.
To this end, we define

\begin{equation}
q(s,s',z,z')=z-z'-v_{h}^{-1}|s-s'|\label{eq:B-07}
\end{equation}
so we have $\partial_{z}q=1$. Then, applying the identity (\ref{eq:B-04}),
we obtain

\begin{equation}
\partial_{z}\Theta\left(q(s,s',z,z')\right)=\delta\left(q(s,s',z,z')\right)\partial_{z}q=\delta\left(q(s,s',z,z')\right)\label{eq:B-08}
\end{equation}
and, substituting Eq. (\ref{eq:B-08}) into Eq. (\ref{eq:B-06}),
we have

\begin{equation}
N_{z}=-\frac{1}{4}v_{h}\int_{-\infty}^{\infty}\left[\int_{-\infty}^{\infty}\delta\left(q(s,s',z,z')\right)\rho(s',z')dz'\right]ds'.\label{eq:B-09}
\end{equation}
Then, by applying identity (\ref{eq:B-05}) to the inner integral
of Eq. (\ref{eq:B-09}), we have

\begin{equation}
\int_{-\infty}^{\infty}\delta\left(q(s,s',z,z')\right)\rho(s',z')dz'=\sum_{i}\frac{\rho(s',r_{i})}{\left|\partial_{z'}q|_{z'=r_{i}}\right|},\label{eq:B-10}
\end{equation}
where, in this case, from Eq. (\ref{eq:B-07}), we have $\partial_{z'}q=-1$
and setting $q(s,s',z,z')=0$ and solving for $z'$ yields one root,
that is 
\begin{equation}
r_{1}=z-v_{h}^{-1}|s-s'|.\label{eq:B-11}
\end{equation}
Therefore, there is only one term in the summation of the RHS of Eq.
(\ref{eq:B-10}). Thus, Eq. (\ref{eq:B-10}) becomes

\begin{equation}
\int_{-\infty}^{\infty}\delta\left(q(s,s',z,z')\right)\rho(s',z')dz'=\frac{\rho(s',r_{1})}{\left|-1\right|}=\rho(s',z-v_{h}^{-1}|s-s'|)\label{eq:B-12}
\end{equation}
and substituting this result into Eq. (\ref{eq:B-09}), we have

\begin{equation}
N_{z}=-\frac{1}{4}v_{h}\int_{-\infty}^{\infty}\rho(s',z-v_{h}^{-1}|s-s'|)ds'.\label{eq:B-13}
\end{equation}
and we see that Eq. (\ref{eq:B-13}) is identical to Eq. (\ref{eq:B-01}).

\subsection{Derivation of $N_{x}$ and $N_{y}$}

\label{subsec:B.2}

Because know $N_{x}=\partial_{x}\phi=\mathrm{sgn}(x)\partial_{s}\phi$
and $N_{y}=\partial_{y}\phi=\mathrm{sgn}(y)\partial_{s}\phi$, we
only need to obtain the correct expression for $\partial_{s}\phi$,
where $\phi$ is again given by Eq. (\ref{eq:A-12}). In other words,
we we differentiate Eq. (\ref{eq:A-12}) w.r.t. $s$, to obtain

\begin{equation}
\partial_{s}\phi=-\frac{1}{4}v_{h}\int_{-\infty}^{\infty}\left[\int_{-\infty}^{\infty}\partial_{s}\Theta\left(z-z'-v_{h}^{-1}|s-s'|\right)\cdot\rho(s',z')ds'\right]dz'.\label{eq:B-14}
\end{equation}
Let us first reverse the order of integration, i.e.

\begin{equation}
\partial_{s}\phi=-\frac{1}{4}v_{h}\int_{-\infty}^{\infty}\left[\int_{-\infty}^{\infty}\partial_{s}\Theta\left(z-z'-v_{h}^{-1}|s-s'|\right)\cdot\rho(s',z')dz'\right]ds'.\label{eq:B-15}
\end{equation}
Again, taking $q$ to be defined in Eq. (\ref{eq:B-07}), we have

\begin{equation}
\partial_{s}q=-v_{h}^{-1}\mathrm{sgn}(s-s').\label{eq:B-16}
\end{equation}
Then, applying identity (\ref{eq:B-04}), we obtain

\begin{equation}
\partial_{s}\Theta\left(q(s,s',z,z')\right)=\delta\left(q(s,s',z,z')\right)\partial_{s}q=-v_{h}^{-1}\mathrm{sgn}(s-s')\delta\left(q(s,s',z,z')\right)\label{eq:B-17}
\end{equation}
and substituting Eq. (\ref{eq:B-17}) into Eq. (\ref{eq:B-15}), we
have

\begin{equation}
\partial_{s}\phi=\frac{1}{4}\int_{-\infty}^{\infty}\mathrm{sgn}(s-s')\left[\int_{-\infty}^{\infty}\delta\left(q(s,s',z,z')\right)\rho(s',z')dz'\right]ds'.\label{eq:B-18}
\end{equation}
Then, applying identity (\ref{eq:B-05}) to the inner integral of
Eq. (\ref{eq:B-18}), we obtain the same result as Eq. (\ref{eq:B-12}),
because $z'$ is - once again - the integration variable. Then, substituting
Eq. (\ref{eq:B-12}) into Eq. (\ref{eq:B-18}), we have

\begin{equation}
\partial_{s}\phi=\frac{1}{4}\int_{-\infty}^{\infty}\mathrm{sgn}(s-s')\rho(s',z-v_{h}^{-1}|s-s'|)ds',\label{eq:B-19}
\end{equation}
and we see that Eq. (\ref{eq:B-19}) is consistent with Eqs. (\ref{eq:B-02})
and (\ref{eq:B-03}).

\section{Analytic evaluation of integrals}

\label{sec:C-evalIntegrals}

This appendix describes the algorithm used to evaluate the integrals
in Eqs. (\ref{eq:B-13}) and (\ref{eq:B-19}), each of which contains
the function\textcolor{green}{{} }$\rho(s',z')$. For the source
functions presented in this paper,\textcolor{green}{{} }$\rho(s',z')$
takes the general form

\begin{equation}
\rho(s{\color{green}'},z{\color{green}'})=W\sum_{i=1}^{4}\alpha_{i}\left[\prod_{\sigma}\Theta\left(p_{\sigma,i}(s{\color{green}'},z{\color{green}'})\right)\right]g_{i}(s{\color{green}'}),\label{eq:C-01}
\end{equation}
where\textcolor{green}{:} 
\begin{itemize}
\item $\Theta$ is the Heaviside function. 
\item The set of $p_{\sigma,i}$ is a family of \textit{shape functions},
where 
\begin{itemize}
\item The index $\sigma$ runs over a number of expressions, which together
determine the source geometry, for example, rhomboids or rectangles. 
\item The index $i$ runs over the centroids $(\beta_{i},\xi_{i})$ of that
geometry. 
\end{itemize}
\item Set $g_{i}$ is a family of \textit{charge profiles} that determines
the charge as a function of the distance from $\beta_{i}$. 
\item The functions $p_{\sigma,i}$ and $g_{i}$ are continuous, but not
necessarily smooth. 
\item $z'=z-v_{h}^{-1}|s-s'|$, as shown in Eqs\textcolor{green}{.} (\ref{eq:B-13})
and (\ref{eq:B-19})\textcolor{green}{.} 
\end{itemize}
The specific shape function(s) and charge profile for each of the
rhomboidal and rectangular sources are presented in Subsections \ref{subsec:C.1}
and \ref{subsec:C.2}.

We will now continue with the general case.\textcolor{green}{{} }From
Eq. (\ref{eq:B-01}) we write

\begin{equation}
\partial_{z}\phi=-\frac{1}{4}v_{h}W\sum_{i=1}^{4}\alpha_{i}\mathcal{I}_{1,i}(s,z)\label{eq:C-02}
\end{equation}
and from Eq. (\ref{eq:B-02}) or (\ref{eq:B-03}) we write

\begin{equation}
\partial_{s}\phi=\frac{1}{4}W\sum_{i=1}^{4}\alpha_{i}\mathcal{I}_{2,i}(s,z),\label{eq:C-03}
\end{equation}
where

\begin{equation}
\mathcal{I}_{1,i}(s,z)=\int_{-\infty}^{\infty}\left[\prod_{\sigma}\Theta\left(p_{\sigma,i}(s',z-v_{h}^{-1}|s-s'|)\right)\right]g_{i}(s')ds'\label{eq:C-04}
\end{equation}
and

\begin{equation}
\mathcal{I}_{2,i}(s,z)=\int_{-\infty}^{\infty}\frac{s-s'}{|s-s'|}\left[\prod_{\sigma}\Theta\left(p_{\sigma,i}(s',z-v_{h}^{-1}|s-s'|)\right)\right]g_{i}(s')ds'.\label{eq:C-05}
\end{equation}
At a fixed point $(s,z)$, the functions $p_{\sigma,i}(s',z-v_{h}^{-1}|s-s'|)$
are piece-wise linear in $s'$, crossing $p_{\sigma,i}(s',z-v_{h}^{-1}|s-s'|)=0$
a finite number of times, where $p_{\sigma,i}(s',z-v_{h}^{-1}|s-s'|)<0$
if $s'$ is less than the smallest root or greater than the largest
root. On the open $s'$-interval between any pair of consecutive roots,
a particular $p_{\sigma,i}(s',z-v_{h}^{-1}|s-s'|)$ is either negative
or positive. If it happens that, for fixed $i$, $p_{\sigma,i}(s',z-v_{h}^{-1}|s-s'|)>0$
for all $\sigma$, then $\prod_{\sigma}\Theta\left(p_{\sigma,i}(s',z-v_{h}^{-1}|s-s'|)\right)=1$;
otherwise $\prod_{\sigma}\Theta\left(p_{\sigma,i}(s',z-v_{h}^{-1}|s-s'|)\right)=0$.
Thus, for each $i$, there exists, at most, a finite number of open
intervals $I_{i,k}^{+}=\left(a_{\sigma_{a},i,k},b_{\sigma_{b},i,k}\right)$\footnote{Note: the subscripts $a$ and $b$, attached to the index $\sigma$,
indicate that $a$ is a $s'$-root of $p_{\sigma_{a},i}(s',z-v_{h}^{-1}|s-s'|)=0$
and $b$ is a $s'$-root of $p_{\sigma_{b},i}(s',z-v_{h}^{-1}|s-s'|)=0$,
where $\sigma_{a}$ may or may not equal $\sigma_{b}$.} that contribute to integrals $\mathcal{I}_{1,i}$ and $\mathcal{I}_{2,i}$.
Based on this analysis, integrals $\mathcal{I}_{1,i}$ and $\mathcal{I}_{2,i}$
can be rewritten as a finite sum of finite integrals (rather than
improper integrals), where the interval bounds become the limits of
integration. That is,

\begin{eqnarray}
\mathcal{I}_{1,i}(s,z) & = & \sum_{k}\int_{a_{\sigma_{a},i,k}}^{b_{\sigma_{b},i,k}}g_{i}(s')ds'\label{eq:C-04a}\\
 & = & \sum_{k}A_{i}(b_{\sigma_{b},i,k})-A_{i}(a_{\sigma_{a},i,k})\label{eq:C-04b}
\end{eqnarray}
and

\begin{eqnarray}
\mathcal{I}_{2,i}(s,z) & = & \sum_{k}\begin{cases}
\int_{a_{\sigma_{a},i,k}}^{b_{\sigma_{b},i,k}}g_{i}(s')ds', & s>b_{\sigma_{b},i,k}\\
-\int_{a_{\sigma_{a},i,k}}^{b_{\sigma_{b},i,k}}g_{i}(s')ds', & s<a_{\sigma_{a},i,k}\\
\int_{a_{\sigma_{a},i,k}}^{s}g_{i}(s')ds'-\int_{s}^{b_{\sigma_{b},i,k}}g_{i}(s')ds', & a_{\sigma_{a},i,k}<s<b_{\sigma_{b},i,k}
\end{cases}\label{eq:C-05a}\\
\nonumber \\ & \!= & \sum_{k}\begin{cases}
A_{i}(b_{\sigma_{b}i,k})-A_{i}(a_{\sigma_{a},i,k}) & s>b_{\sigma_{b},i,k}\\
A_{i}(a_{\sigma_{a},i,k})-A_{i}(b_{\sigma_{b},i,k}), & s<a_{\sigma_{a},i,k}\\
2A_{i}(s)-A_{i}(a_{\sigma_{a},i,k})-A_{i}(b_{\sigma_{b},i,k}), & a_{\sigma_{a},i,k}<s<b_{\sigma_{b},i,k}
\end{cases},\label{eq:C-05b}
\end{eqnarray}
respectively, where: 
\begin{itemize}
\item The quantity$A_{i}$ denotes an antiderivative of $g_{i}$. 
\item The factor $\frac{s-s'}{|s-s'|}$, which is expressed in Eq. (\ref{eq:C-05})
was eliminated by breaking each interval into three parts, as shown
in Eqs. (\ref{eq:C-05a}) and (\ref{eq:C-05b}), respectively.
\end{itemize}
The intervals $I_{i,k}^{+}$ are determined by the following algorithm: 
\begin{enumerate}
\item Obtain the set of $s'$-roots $\left\{ r_{\sigma,i,j},\,j=1,2,...\right\} $
of $p_{\sigma,i}(s',z-v_{h}^{-1}|s-s'|)=0$, for each $\sigma$. 
\item Obtain the union $\left\{ r_{i,j}\right\} =\cup_{\sigma}$$\left\{ r_{\sigma,i,j}\right\} $
of all roots. 
\item Sort the set $\left\{ r_{i,j}\right\} $ into ascending order, and
define the set of consecutive, open intervals $\left\{ I_{i,j}=\left(r_{i,j},r_{i,j+1}\right)\right\} $. 
\item Obtain the subset of intervals $\left\{ I_{i,k}^{+}\right\} \subset\left\{ I_{i,j}\right\} $,
such that $p_{\sigma,i}(\mu_{i,j},z-v_{h}^{-1}|s-\mu_{i,j}|)>0$ for
all $\sigma$, where $\mu_{i,j}=(r_{i,j}+r_{i,j+1})/2$ is the midpoint
of the interval $I_{i,j}$. 
\end{enumerate}

\subsection{Specifics for rhomboidal sources}

\label{subsec:C.1}

For the rhomboidal source function, the set $\left\{ p_{\sigma}\right\} $\textcolor{green}{{}
}has only one member; therefore, we drop the index $\sigma$. This
single shape function is

\begin{equation}
p_{i}(s',z-v_{h}^{-1}|s-s'|)=L_{x}-\frac{L_{x}}{L_{z}}\left|z-\xi_{i}-\frac{1}{v_{h}}|s-s'|\right|-\left||s'|-\beta_{i}\right|\label{eq:C-08}
\end{equation}
and the corresponding charge profile is

\begin{equation}
g_{i}(s')=Q_{\text{min}}+\left(Q_{\text{max}}-Q_{\text{min}}\right)\left(|s'|-\beta_{i}\right)^{2},\label{eq:C-09}
\end{equation}
where $p_{i}$ and $g_{i}$ are given by Eqs. (\ref{Eq:shapeRhom})
and (\ref{Eq:chgPflRhom}), respectively, and\textcolor{green}{{} }the
various parameters are described in Section (\ref{subsec:de}).

Now, given Eqs. (\ref{eq:C-08}) and (\ref{eq:C-09}), the geometry-dependent
details of the algorithm are as follows:
\begin{enumerate}
\item Obtain the set of roots $\left\{ r_{i,j}\right\} $ of $p_{i}(s',z-v_{h}^{-1}|s-s'|)=0$
by equating the RHS of Eq. (\ref{eq:C-08}) to zero and solving for
$s'$. This entails the tedious process of unfolding pairs of nested
absolute values - a process that yields 16 equations. These 16 equations
are summarized as by
\begin{equation}
s'_{i,j}=\frac{T_{j,1}L_{x}+T_{j,2}\beta_{i}+T_{j,3}\lambda\left(z-\xi_{i}\right)+T_{j,4}\kappa\lambda s}{1+T_{j,4}\kappa\lambda}\label{eq:C-10}
\end{equation}
where $\lambda=L_{x}/L_{z}$, $\kappa=v_{h}^{-1}$ and $T_{j,\nu},\:j=1,2,...,16,\:\nu=1,2,3,4$
are elements of a $16\times4$ table of signs, with each of the 16
rows being a unique combination of four signs. However, only a subset
of $\left\{ s'_{i,j}\right\} $ actually satisfies $p_{i}(s',z-v_{h}^{-1}|s-s'|)=0$.
To determine this subset, we substituted each member of $\left\{ s'_{i,j}\right\} $
into Eq. (\ref{eq:C-08}), to test whether the RHS is zero. The set
of roots $\left\{ r_{i,j}\right\} $ that we seek is the subset of
the $\left\{ s'_{i,j}\right\} $ whose members pass this test, i.e.
$\left\{ r_{i,j}\right\} \subset\left\{ s'_{i,j}:p_{i}(s',z-v_{h}^{-1}|s-s'|)=0\right\} $. 
\item Because there is only one function $p_{i}$, the union over $\sigma$
is just $\left\{ r_{i,j}\right\} $. 
\end{enumerate}
After performing generic steps 3 and 4, we obtain the set of intervals
$I_{k}^{+}=\left(a_{i,k},b_{i,k}\right)$ for the rhomboidal source
function. Also, from Eq. (\ref{eq:C-09}), we obtain

\begin{equation}
A_{i}(s')=Q_{\min}s'+\left(Q_{\max}-Q_{\min}\right)\left(\frac{\left(s'\right)^{3}}{3}-\beta_{i}s'|s'|+\beta_{i}^{2}s'\right)\label{eq:C-11}
\end{equation}
which we used to compute $A_{i}(a_{i,k})$ and $A_{i}(b_{i,k})$ for
all $k$, which we then substituted into Eqs. (\ref{eq:C-04b}) and
(\ref{eq:C-05b}), respectively.

\subsection{Specifics for rectangular sources}

\label{subsec:C.2}

For the rectangular source function, the set $\left\{ p_{\sigma}\right\} $
has two members are

\begin{equation}
p_{1,i}(s',z-v_{h}^{-1}|s-s'|)=L_{z}-\left|z-\xi_{i}-\frac{1}{v_{h}}|s-s'|\right|\label{eq:C-12}
\end{equation}
and

\begin{equation}
p_{2,i}(s',z-v_{h}^{-1}|s-s'|)=L_{x}-\left||s'|-\beta_{i}\right|\label{eq:C-13}
\end{equation}
and the corresponding charge profile is

\begin{equation}
g_{i}(s')=Q_{\max}.\label{eq:C-14}
\end{equation}
where $p_{1,i}$ and $p_{2,i}$ are given by the first two lines in
Eq. (\ref{sec:SrcRect})\textcolor{green}{, }and where $g_{i}(s')$
is given by the third line in Eq. (\ref{sec:SrcRect})\textcolor{green}{.}

Now, given Eqs. (\ref{eq:C-12}) and (\ref{eq:C-13}), the geometry-dependent
details of the algorithm are as follows: 
\begin{enumerate}
\item Obtain the sets of roots $\left\{ r_{\sigma,i,j}\right\} $ of $p_{\sigma,i}(s',z-v_{h}^{-1}|s-s'|)=0$,
for $\sigma=1,2$, by equating the RHS of both Eq. (\ref{eq:C-12})
and Eq. (\ref{eq:C-13}) to zero, and solving each equation for $s'$.
This entails unfolding the nested absolute value in each equation,
a process that yields four equations for each. It can be shown that
these two sets of four equations are summarized by 
\begin{equation}
s'_{1,i,j}=\frac{T_{j,1}L_{z}+T_{j,2}(z-\xi_{i})+\kappa s}{\kappa}\label{eq:C-15}
\end{equation}
and 
\begin{equation}
s'_{2,i,j}=T_{j,1}L_{x}+T_{j,2}\beta_{i}\label{eq:C-16}
\end{equation}
where $T_{j,\nu},\:j=1,2,3,4,\:\nu=1,2$ are elements of a $4\times2$
table of signs, with each of the four rows being a unique combination
of two signs. However, only a subset of each set $\left\{ s_{\sigma,i,j}\right\} $
satisfies the corresponding equation $p_{\sigma,i}(s',z-v_{h}^{-1}|s-s'|)=0$.
To determine these subsets, we substituted each member of $\left\{ s'_{1,i,j}\right\} $
and $\left\{ s'_{2,i,j}\right\} $ into Eqs. (\ref{eq:C-12}) and
(\ref{eq:C-13}), respectively, to test whether the RHS of each is
zero. The two sets of roots $\left\{ r_{2,i,j}\right\} $ and $\left\{ r_{2,i,j}\right\} $
that we seek are the subsets of $\left\{ s'_{1,i,j}\right\} $ and
$\left\{ s'_{2,i,j}\right\} $, whose members pass their respective
tests, i.e. $\left\{ r_{\sigma,i,j}\right\} \subset\left\{ s'_{\sigma,i,j}:p_{\sigma,i}(s',z-v_{h}^{-1}|s-s'|)=0\right\} $\textcolor{green}{{}
}for $\sigma=1,2$. 
\item Because there are two function $p_{1,i}$ and $p_{2,i}$, we have
$\left\{ r_{i,j}\right\} =\cup_{\sigma}\left\{ r_{\sigma,i,j}\right\} =$$\left\{ r_{1,i,j}\right\} \cup\left\{ r_{2,i,j}\right\} $. 
\end{enumerate}
After performing generic steps 3 and 4, we obtain the set of intervals
$I_{k}^{+}=\left(a_{\sigma_{a}i,k},b_{\sigma_{b}i,k}\right)$ for
the rectangular source function. Note that, for some intervals, we
may have $\sigma_{a}=\sigma_{b}$; that is, the bounds are the roots
of the same function. Also, from Eq. (\ref{eq:C-14}), we obtain

\begin{equation}
A_{i}(s')=Q_{\max}s'\label{eq:C-17}
\end{equation}
which we use to compute $A_{i}(a_{i,k})$ and $A_{i}(b_{i,k})$ for
all $k$, which we then substitute into Eqs. (\ref{eq:C-04b}) and
(\ref{eq:C-05b}), respectively.

\section{Partial Derivatives of the Shift Vector}

\label{sec:D-derivatives}

In this appendix, we derive analytic expressions for the derivatives
of the shift vector components. It is convenient to write two equations:
one for the three partial derivatives of $N_{z}$ and one for the
three partial derivatives of both $N_{x}$ and $N_{y}$. We write
these as follows:

\begin{equation}
\partial_{u}N_{z}(s,z)=-\frac{1}{4}v_{h}W\sum_{i=1}^{4}\alpha_{i}\partial_{u}\mathcal{I}_{1,i}(s,z)\label{eq:duNz}
\end{equation}
and

\begin{equation}
\partial_{u}N_{v}(s,z)=\frac{1}{4}W\frac{v}{|v|}\sum_{i=1}^{4}\alpha_{i}\partial_{u}\mathcal{I}_{2,i}(s,z),\label{eq:duNv}
\end{equation}
where $u\in\left\{ x,y,z\right\} $, $v\in\left\{ x,y\right\} $ and
$\mathcal{I}_{1,i}$ and $\mathcal{I}_{2,i}$ are given by Eqs. (\ref{eq:C-04a})
and (\ref{eq:C-05a}), respectively.

\subsection{Partial derivatives of the integrals $\mathcal{I}_{1,i}$ and $\mathcal{I}_{2,i}$}

\label{subsec:duI1,2}

The first step in evaluating Eqs. (\ref{eq:duNz}) and (\ref{eq:duNv})
is to obtain expressions for $\partial_{u}\mathcal{I}_{1,i}$ and
$\partial_{u}\mathcal{I}_{2,i}$. Expanding the derivatives in Eqs.
(\ref{eq:duNz}) and (\ref{eq:duNv}), by substituting RHS of Eqs.
(\ref{eq:C-04a}) and (\ref{eq:C-05a}), respectively, we obtain

\begin{equation}
\partial_{u}\mathcal{I}_{1,i}(s,z)=\sum_{k}\partial_{u}\int_{a_{\sigma_{a},i,k}}^{b_{\sigma_{b},i,k}}g_{i}(s')ds'\label{eq:duI1}
\end{equation}
and

\begin{equation}
\partial_{u}\mathcal{I}_{2,i}(s,z)=\sum_{k}\begin{cases}
\partial_{u}\int_{a_{\sigma_{a},i,k}}^{b_{\sigma_{b},i,k}}g_{i}(s')ds', & s>b_{\sigma_{b},i,k}\\
-\partial_{u}\int_{a_{\sigma_{a},i,k}}^{b_{\sigma_{b},i,k}}g_{i}(s')ds', & s<a_{\sigma_{a},i,k}\\
\partial_{u}\left(\int_{a_{\sigma_{a},i,k}}^{s}g_{i}(s')ds'-\int_{s}^{b_{\sigma_{b},i,k}}g_{i}(s')ds'\right), & a_{\sigma_{a},i,k}<s<b_{\sigma_{b},i,k}
\end{cases}\label{eq:duI2}
\end{equation}
Comparing Eq. (\ref{eq:duI1}) with Eq. (\ref{eq:duI2}), it can be
observed that: 
\begin{itemize}
\item The first line of the latter is identical to the former; that is,
if $s>b_{\sigma_{b},i,k}$, then $\partial_{u}\mathcal{I}_{2,i}=\partial_{u}\mathcal{I}_{1,i}$. 
\item The second line of the latter is the negative of the former; that
is, if $s<a_{\sigma_{a},i,k}$, then $\partial_{u}\mathcal{I}_{2,i}=-\partial_{u}\mathcal{I}_{1,i}$. 
\end{itemize}
Using these facts, we may rewrite Eq. (\ref{eq:duI2}) as

\begin{equation}
\partial_{u}\mathcal{I}_{2,i}(s,z)=\sum_{k}\begin{cases}
\partial_{u}\mathcal{I}_{1,i}(s,z), & s>b_{\sigma_{b},i,k}\\
-\partial_{u}\mathcal{I}_{1,i}(s,z), & s<a_{\sigma_{a},i,k}\\
\partial_{u}\left(\int_{a_{\sigma_{a},i,k}}^{s}g_{i}(s')ds'-\int_{s}^{b_{\sigma_{b},i,k}}g_{i}(s')ds'\right), & a_{\sigma_{a},i,k}<s<b_{\sigma_{b},i,k}
\end{cases}\label{eq:duI2a}
\end{equation}
Thus, we can see that there are only two unique integrations to be
evaluated. Recalling that the limits of integration $a_{\sigma_{1},i,k}$
and $b_{\sigma_{2},i,k}$ are the roots of equations $p_{\sigma,i}\left(s',z-v_{h}^{-1}|s-s'|\right)=0$,
and thus depend on the coordinates $(s,z)$, we use the standard result
for differentiating a definite integral with variable limits. Doing
this for Eq. (\ref{eq:duI1}) and the third line in Eq. (\ref{eq:duI2a}),
we obtain

\begin{equation}
\partial_{u}\mathcal{I}_{1,i}(s,z)=\sum_{k}\left(g_{i}(b_{\sigma_{b},i,k})\partial_{u}b_{\sigma_{b},i,k}-g_{i}(a_{\sigma_{a},i,k})\partial_{u}a_{\sigma_{a},i,k}\right)\label{eq:duI1a}
\end{equation}
and (after a bit of algebra)

\begin{equation}
\partial_{u}\mathcal{I}_{2,i}(s,z)=\sum_{k}\left(2g_{i}(s)\partial_{u}s-g_{i}(a_{\sigma_{a},i,k})\partial_{u}a_{\sigma_{a},i,k}-g_{i}(b_{\sigma_{b},i,k})\partial_{u}b_{\sigma_{b},i,k}\right)\label{eq:duI2b}
\end{equation}
respectively.

Substituting Eqs. (\ref{eq:duI1a}) and (\ref{eq:duI2b}) into (\ref{eq:duNz})
and (\ref{eq:duNv}), respectively, we obtain

\begin{equation}
\partial_{u}N_{z}(s,z)=-\frac{1}{4}v_{h}W\sum_{i=1}^{4}\alpha_{i}\sum_{k}\left(g_{i}(b_{\sigma_{b},i,k})\partial_{u}b_{\sigma_{b},i,k}-g_{i}(a_{\sigma_{a},i,k})\partial_{u}a_{\sigma_{a},i,k}\right)\label{eq:duNz-a}
\end{equation}
and

\begin{equation}
\partial_{u}N_{v}(s,z)=\frac{1}{4}W\frac{v}{|v|}\sum_{i=1}^{4}\alpha_{i}\sum_{k}\left(2g_{i}(s)\partial_{u}s-g_{i}(a_{\sigma_{a},i,k})\partial_{u}a_{\sigma_{a},i,k}-g_{i}(b_{\sigma_{b},i,k})\partial_{u}b_{\sigma_{b},i,k}\right).\label{eq:duNv-a}
\end{equation}

\subsection{Partial derivatives of the roots $a$ and $b$}

\label{subsec:dur}

As shown in Eqs. (\ref{eq:duI1a}) and (\ref{eq:duI2b}), we require
the derivatives of the roots $a_{\sigma_{1},i,k}$ and $b_{\sigma_{2},i,k}$
with respect to the coordinates $u\in\left\{ s,z\right\} $. Considering
this, we define

\begin{equation}
P_{\sigma,i}(s,z)=p_{\sigma,i}\left(s',z-v_{h}^{-1}|s-s'|\right)\label{eq:P}
\end{equation}
where for now $s'$ is an arbitrary function of $s$ and $z$. The
total derivative of $P_{\sigma,i}$ is

\begin{equation}
\frac{dP_{\sigma,i}}{du}=\frac{\partial p_{\sigma,i}}{\partial s'}\frac{\partial s'}{\partial u}+\frac{\partial p_{\sigma,i}}{\partial u}.
\end{equation}
Let $s'(s,z)$ be defined\footnote{$s'$ is not uniquely defined, because there are multiple roots for
each value of $s$ and $z$. However, we will take one of those roots
$r_{0}$ for a particular $s_{0}$ and $z_{0}$ and then define $s'$
for points in the neighborhood of $s_{0}$ and $z_{0}$ to be the
roots that are close to $r_{0}$. This operation can be made mathematically
rigorous, resulting in a differentiable function $s'(s,z)$. This
procedure applies for each root.} by $p_{\sigma,i}\left(s',z-v_{h}^{-1}|s-s'|\right)=0$. Then, by
using the definition in Eq. (\ref{eq:P}), $P_{\sigma,i}(s,z)=p_{\sigma,i}\left(s',z-v_{h}^{-1}|s-s'|\right)=0$,
so

\begin{equation}
\frac{\partial p_{\sigma,i}}{\partial s'}\frac{\partial s'}{\partial u}+\frac{\partial p_{\sigma,i}}{\partial u}=0.\label{eq:P2}
\end{equation}
For clarity, we switch to index notation, wherein Eq. (\ref{eq:P2}),
can be written as

\begin{equation}
\partial_{s'}p_{\sigma,i}\partial_{u}s'+\partial_{u}p_{\sigma,i}=0.\label{eq:implDiff}
\end{equation}
Then, solving Eq. \ref{eq:implDiff} for $\partial_{u}s'$, we have

\begin{equation}
\partial_{u}s'=-\frac{\partial_{u}p_{\sigma,i}(s',z-v_{h}^{-1}|s-s'|)}{\partial_{s'}p_{\sigma,i}(s',z-v_{h}^{-1}|s-s'|)}.\label{eq:dus'}
\end{equation}
Then, denoting roots $a_{\sigma_{a},i,k}$ and $b_{\sigma_{b},i,k}$
collectively by $r_{\sigma,i,k}$, we have

\begin{equation}
\partial_{u}r_{\sigma,i,k}=\partial_{u}s'|_{s'=r_{\sigma,i,k}}=-\frac{\partial_{u}p_{\sigma,i}(s',z-v_{h}^{-1}|s-s'|)|_{s'=r_{\sigma,i,k}}}{\partial_{s'}p_{\sigma,i}(s',z-v_{h}^{-1}|s-s'|)|_{s'=r_{\sigma,i,k}}}\coloneqq-q_{\sigma,i,k}^{u}(r_{\sigma,i,k}),\label{eq:dur}
\end{equation}
where for convenience, we denote the quotient appearing in Eq. (\ref{eq:dur})
by $q_{\sigma,i,k}^{u}$.

Now, recall that Eq. (\ref{eq:dur}) represents three equations, due
to $u\in\{x,y,z\}$. For clarity, we explicitly wrote these out (see
Eqs. (\ref{eq:dzr}), (\ref{eq:dxr}) and (\ref{eq:dyr}), below).
Additionally, for brevity, we suppress the arguments to the $p_{\sigma,i}$
(and bear in mind that the $p_{\sigma,i}$ are, in general, functions
of $s'$, $s$ and $z$). For $u=z$, we have

\begin{equation}
\partial_{z}r_{\sigma,i,k}=\partial_{z}s'|_{s'=r_{\sigma,i,k}}=-\frac{\partial_{z}p_{\sigma,i}|_{s'=r_{\sigma,i,k}}}{\partial_{s'}p_{\sigma,i}|_{s'=r_{\sigma,i,k}}}\coloneqq-q_{\sigma,i,k}^{z}(r_{\sigma,i,k})\label{eq:dzr}
\end{equation}
and, since $s=|x|+|y|$, for $u\in\{x,y\}$, we have

\begin{equation}
\partial_{x}r_{\sigma,i,k}=-\frac{\partial_{s}p_{\sigma,i}|_{s'=r_{\sigma,i,k}}}{\partial_{s'}p_{\sigma,{\color{green}i}}|_{s'=r_{\sigma,i,k}}}\partial_{x}s\coloneqq-q_{\sigma,i,k}^{s}(r_{\sigma,i,k})\partial_{x}s=-\frac{x}{|x|}q_{\sigma,i,k}^{s}(r_{\sigma,i,k})\label{eq:dxr}
\end{equation}
and

\begin{equation}
\partial_{y}r_{\sigma,i,k}=-\frac{\partial_{s}p_{\sigma,i}|_{s'=r_{\sigma,i,k}}}{\partial_{s'}p_{\sigma,{\color{green}i}}|_{s'=r_{\sigma,i,k}}}\partial_{y}s\coloneqq-q_{\sigma,i,k}^{s}(r_{\sigma,i,k})\partial_{y}s=-\frac{y}{|y|}q_{\sigma,i,k}^{s}(r_{\sigma,i,k}).\label{eq:dyr}
\end{equation}

Next, substituting Eq. (\ref{eq:dur}) into Eqs. (\ref{eq:duNz-a})
and (\ref{eq:duNv-a}), we obtain, respectively

\begin{equation}
\partial_{u}N_{z}(s,z)=-\frac{1}{4}v_{h}W\sum_{i=1}^{4}\alpha_{i}\Omega_{1,i}^{u}\label{eq:duNz-b}
\end{equation}
and

\begin{eqnarray}
\partial_{u}N_{v}(s,z) & = & \frac{1}{4}W\frac{v}{|v|}\sum_{i=1}^{4}\alpha_{i}\Omega_{2,i}^{u}.\label{eq:duNv-b}
\end{eqnarray}
where we have defined

\begin{equation}
\Omega_{1,i}^{u}=\sum_{k}\left(g_{i}(b_{\sigma_{b},i,k})q_{\sigma_{b},i,k}^{u}(b_{\sigma,i,k})-g_{i}(a_{\sigma_{a},i,k})q_{\sigma_{a},i,k}^{u}(a_{\sigma,i,k})\right)\label{eq:Omega1}
\end{equation}
and

\begin{equation}
\Omega_{2,i}^{u}=\sum_{k}\left(2g_{i}(s)\partial_{u}s+g_{i}(a_{\sigma_{a},i,k})q_{\sigma_{a},i,k}^{u}(b_{\sigma,i,k})+g_{i}(b_{\sigma_{b},i,k})q_{\sigma_{b},i,k}^{u}(b_{\sigma,i,k})\right)\label{eq:Omega2}
\end{equation}
Each of the Eqs. (\ref{eq:duNz-b}) and (\ref{eq:duNv-b}) represent
the next three equations\textcolor{green}{.} That is, for $u=z,x,y$,
we have

\begin{eqnarray}
\partial_{z}N_{z}(s,z) & = & -\frac{1}{4}v_{h}W\sum_{i=1}^{4}\alpha_{i}\Omega_{1,i}^{z}\label{eq:dzNz}\\
\partial_{x}N_{z}(s,z) & = & -\frac{1}{4}v_{h}W\frac{x}{|x|}\sum_{i=1}^{4}\alpha_{i}\Omega_{1,i}^{x}\label{eq:dxNz}\\
\partial_{y}N_{z}(s,z) & = & -\frac{1}{4}v_{h}W\frac{y}{|y|}\sum_{i=1}^{4}\alpha_{i}\Omega_{1,i}^{y}\label{eq:dyNz}
\end{eqnarray}
and for $u\in\{x,y\}$, we have, respectively

\begin{eqnarray}
\partial_{x}N_{x}(s,z) & = & \frac{1}{4}W\frac{x}{|x|}\frac{x}{|x|}\sum_{i=1}^{4}\alpha_{i}\Omega_{2,i}^{x},\quad v=x\label{eq:dxNx}\\
\partial_{x}N_{y}(s,z) & = & \frac{1}{4}W\frac{x}{|x|}\frac{y}{|y|}\sum_{i=1}^{4}\alpha_{i}\Omega_{2,i}^{y},\quad v=y\label{eq:dxNy}\\
\partial_{y}N_{y}(s,z) & = & \frac{1}{4}W\frac{y}{|y|}\frac{y}{|y|}\sum_{i=1}^{4}\alpha_{i}\Omega_{2,i}^{y},\quad v=y\label{eq:dyNy}
\end{eqnarray}
and of course $\partial_{y}N_{x}=\partial_{x}N_{y}$, $\partial_{z}N_{x}=\partial_{x}N_{z}$
and $\partial_{y}N_{z}=\partial_{z}N_{x}$, accounting for all nine
partial derivatives of the shift vector.

\subsection{Partial derivatives of the shape functions}

\label{subsec:dup}

We now require expressions for the derivatives $\partial_{z}p_{\sigma,i}$
and $\partial_{s}p_{\sigma,i}$, which appear in the numerators of
$q^{z}$ and $q^{s}$, respectively, and $\partial_{s'}p_{\sigma,i}$,
which appears in both denominators. These are provided for rhomboidal
and rectangular shape functions in Sections \ref{subsec:dpRhom} and
\ref{subsec:dpRect}, respectively.

\subsubsection{Partial derivatives of the rhomboidal shape function}

\label{subsec:dpRhom}

By taking the partial derivatives of Eq. \ref{eq:C-08} wrt $z$,
$s$ and $s'$, we obtain for any root $r$

\begin{equation}
\partial_{z}p_{i}(s',z-v_{h}^{-1}|s-s'|)|_{s'=r}=-\frac{L_{x}}{L_{z}}\frac{\eta(r)}{\left|\eta(r)\right|},\quad\eta(r)\neq0\label{eq:dzp}
\end{equation}
where

\begin{equation}
\eta(r)=z-\xi_{i}-\frac{1}{v_{h}}|s-r|,\label{eq:eta}
\end{equation}

\begin{equation}
\partial_{s}p_{i}(s',z-v_{h}^{-1}|s-s'|)|_{s'=r}=\frac{1}{v_{h}}\frac{L_{x}}{L_{z}}\frac{s-r}{|s-r|}\frac{\eta(r)}{\left|\eta(r)\right|},\quad r\neq s,\eta(r)\neq0\label{eq:dsp}
\end{equation}
and

\begin{equation}
\partial_{s'}p_{i}(s',z-v_{h}^{-1}|s-s'|)|_{s'=r}=-\frac{r}{\left|r\right|}\frac{|r|-\beta_{i}}{\left||r|-\beta_{i}\right|}-\frac{1}{v_{h}}\frac{L_{x}}{L_{z}}\frac{s-r}{\left|s-r\right|}\frac{\eta(r)}{\left|\eta(r)\right|},\quad r\neq0,|r|\neq\beta_{i},\eta(r)\neq0.\label{eq:ds'p}
\end{equation}

\subsubsection{Partial derivatives of the rectangular shape functions}

\label{subsec:dpRect}

Here we have two shape functions, $p_{\sigma}$, $\sigma=1,2$, whose
derivatives we treat separately.

\paragraph{Partial derivatives of $p_{1}$\protect \protect \protect \protect \\
 }

By taking the partial derivatives of Eq. (\ref{eq:C-12}) wrt $z$,
$s$ and $s'$, we obtain for any root $r$

\begin{eqnarray}
\partial_{z}p_{1,i}(s',z-v_{h}^{-1}|s-s'|)|_{s'=r} & = & -\frac{\eta(r)}{\left|\eta(r)\right|},\quad\eta(r)\neq0\label{eq:dzp1}
\end{eqnarray}
where $\eta(r)$ is given by Eq. (\ref{eq:eta}),

\begin{eqnarray}
\partial_{s}p_{1,i}(s',z-v_{h}^{-1}|s-s'|)|_{s'=r} & = & \frac{1}{v_{h}}\frac{s-r}{|s-r|}\frac{\eta(r)}{\left|\eta(r)\right|},\quad r\neq s,\eta(r)\neq0\label{eq:dsp1}
\end{eqnarray}
and

\begin{eqnarray}
\partial_{s'}p_{1,i}(s',z-v_{h}^{-1}|s-s'|)|_{s'=r} & = & -\frac{1}{v_{h}}\frac{s-r}{|s-r|}\frac{\eta(r)}{\left|\eta(r)\right|},\quad r\neq s,\eta(r)\neq0.\label{eq:ds'p1}
\end{eqnarray}

\paragraph{Partial derivatives of $p_{2}$\protect \protect \protect \protect \\
 }

By taking the partial derivatives of Eq. (\ref{eq:C-13}) wrt $z$,
$s$ and $s'$, we obtain

\begin{equation}
\partial_{z}p_{2,i}(s',z-v_{h}^{-1}|s-s'|)|_{s'=r}=0,\label{eq:dzp2}
\end{equation}

\begin{equation}
\partial_{s}p_{2,i}(s',z-v_{h}^{-1}|s-s'|)|_{s'=r}=0\label{eq:dsp2}
\end{equation}
and

\begin{eqnarray}
\partial_{s'}p_{2,i}(s',z-v_{h}^{-1}|s-s'|)|_{s'=r} & = & -\frac{r}{|r|}\frac{|r|-\beta_{i}}{\left||r|-\beta_{i}\right|},\quad r\neq0,|r|\neq\beta_{i}.\label{eq:ds'p2}
\end{eqnarray}

\subsection{Extensions}

\label{subsec:Ext}

From the above equations, we see that the partial derivatives of the
shift vector are undefined under various circumstances. We refer to
the domains of undefinition as ``singular regions.'' It is clear
that our theory needs to be modified slightly so that these quantities
are no longer undefined. However, different modifications generally
lead to different values of those partial derivatives.\textcolor{green}{{}
}Various principles can be invoked to limit the ambiguities. Most
importantly, we would like physically measurable quantities to be
insensitive to modification. For example, we require that integrals
of the energy density are independent of these modifications. This
is easily achieved when the singular regions lie along a set of measure
0 (in 3D). In that case, we can extend our functions such that, along
these regions, the assigned values are finite. Any finite assignment
results in identical integrated values. We believe that with these
finite assignments, other physically measurable quantities will also
be independent of the values of their values. In the tables that follow,
we indicate the particular circumstances and the equations that indicate
those circumstances, for which each of the above-mentioned partial
derivatives are undefined. We then go on to state the finite values
we have used to extend our functions into singular regions.\\

Circumstances under which partials of the shift vector are undefined:

\begin{tabular}{|c|c|c|}
\hline 
Partial of shift vector  & Eq.  & Circumstances\tabularnewline
\hline 
\hline 
$\partial_{z}N_{z}$ undefined  & (\ref{eq:dzNz})  & $q^{z}$ undefined\tabularnewline
\hline 
$\partial_{x}N_{z}$ undefined  & (\ref{eq:dxNz})  & $x=0$ or $q^{s}$ undefined\tabularnewline
\hline 
$\partial_{y}N_{z}$ undefined  & (\ref{eq:dyNz})  & $y=0$ or $q^{s}$ undefined\tabularnewline
\hline 
$\partial_{x}N_{x}$ undefined  & (\ref{eq:dxNx})  & $x=0$ or $q^{s}$ undefined\tabularnewline
\hline 
$\partial_{y}N_{y}$ undefined  & (\ref{eq:dyNy})  & $y=0$ or $q^{s}$ undefined\tabularnewline
\hline 
$\partial_{x}N_{y}$ undefined  & (\ref{eq:dxNy})  & $x=0$, $y=0$ or $q^{s}$ undefined\tabularnewline
\hline 
\end{tabular}\\

Circumstances under which quotients $q^{z}$ and $q^{s}$ are undefined:

\begin{tabular}{|c|c|c|}
\hline 
Quotient  & Eq.  & Circumstances\tabularnewline
\hline 
\hline 
$q^{z}$ undefined  & (\ref{eq:dur})  & $\partial_{z}p_{\sigma,i}$ or $\partial_{s'}p_{\sigma,i}$ undefined\tabularnewline
\hline 
$q^{s}$undefined  & (\ref{eq:dxr})  & $\partial_{s}p_{\sigma,i}$ or $\partial_{s'}p_{\sigma,i}$ undefined\tabularnewline
\hline 
\end{tabular}\\

Circumstances under which partials of the rhomboidal shape function
is undefined:

\begin{tabular}{|c|c|c|}
\hline 
Partial of shape func  & Eq.  & Circumstances\tabularnewline
\hline 
\hline 
$\partial_{z}p_{i}$ undefined  & (\ref{eq:dzp})  & $\eta(r)=0$\tabularnewline
\hline 
$\partial_{s}p_{i}$ undefined  & (\ref{eq:dsp})  & $r=s$ or $\eta(r)=0$\tabularnewline
\hline 
$\partial_{s'}p_{i}$ undefined  & (\ref{eq:ds'p})  & $r=0$, $|r|=\beta_{i}$, $r=s$ or $\eta(r)=0$\tabularnewline
\hline 
\end{tabular}\\

Circumstances under which partials of the rectangular shape $p_{1}$
are undefined:

\begin{tabular}{|c|c|c|}
\hline 
Partial of shape func  & Eq.  & Circumstances\tabularnewline
\hline 
\hline 
$\partial_{z}p_{1,i}$ undefined  & (\ref{eq:dzp1})  & $\eta(r)=0$\tabularnewline
\hline 
$\partial_{s}p_{1,i}$ undefined  & (\ref{eq:dsp1})  & $r=s$ or $\eta(r)=0$\tabularnewline
\hline 
$\partial_{s'}p_{1,i}$ undefined  & (\ref{eq:ds'p1})  & $r=s$ or $\eta(r)=0$\tabularnewline
\hline 
\end{tabular}\\

Circumstances under which partials of the rectangular shape $p_{2}$
are undefined:

\begin{tabular}{|c|c|c|}
\hline 
Partial of shape func  & Eq.  & Circumstances\tabularnewline
\hline 
\hline 
$\partial_{z}p_{2,i}$ undefined  & (\ref{eq:dzp1})  & None\tabularnewline
\hline 
$\partial_{s}p_{2,i}$ undefined  & (\ref{eq:dsp1})  & None\tabularnewline
\hline 
$\partial_{s'}p_{2,i}$ undefined  & (\ref{eq:ds'p1})  & $r=0$ or $|r|=\beta_{i}$\tabularnewline
\hline 
\end{tabular}\\
 \\
 Ultimately, all the above circumstances occur at points where expressions
for the partials of the shift vector or of the shape function contain
a factor of the form $w/|w|$, which is undefined at $w=0$. At such
points, reasonable values are assigned, thereby extending (or completing)
the expressions for the derivatives at the points in question.

For mixed derivatives of the shift vector, that is, $\partial_{u}N_{v},\;u\neq v$,
there are overall factors $u/|u|$, $u\in\{x,y\}$. A reasonable extension
for these points consists of substituting the signum function $\mathrm{sgn}(u)$
for all ratios $u/|u|$, with the usual convention that $\mathrm{sgn}(0)=0$.
Because $\mathrm{sgn}(u)$ is equivalent to the original ratio everywhere
except at $u=0$, we merely substitute zero at $u=0$. On the other
hand, for pure derivatives of the shift vector, i.e. $\partial_{u}N_{u}$,
the substitution produces the factor $\mathrm{sgn}^{2}(u)$. As this
has a value of 1 for all $u\neq0$, it is reasonable to set $\mathrm{sgn}^{2}(u)=1$
at $u=0$.

For derivatives of the shape function, we assign different extensions
depending on whether we are dealing with the rhomboidal or rectangular
case, as discussed in Sections \ref{subsec:ExtRhom} and\ref{subsec:ExtRect},
respectively.

\subsubsection{Extensions for the rhomboidal case}

\label{subsec:ExtRhom}

Here, all the undefined points result from expressions of the form
$w/|w|$. Following the same reasoning as above, all extensions for
the rhomboidal case involve substituting the signum function $\mathrm{sgn}(w)$
for all ratios $w/|w|$. There is one possible objection to this choice,
which we will now lay to rest. Note that $\partial_{s'}p$ appears
in the denominator of the quotients, as shown in Eq. (\ref{eq:dur}),
and for rhomboidal case, $\partial_{s'}p$ is given by Eq. (\ref{eq:ds'p}).
According to the latter, two terms depend on the ratios of the form
$w/|w|$. Thus, if one such ratio in each term is set to zero, then
the quotients are undefined. Specifically, if either $r=0$ or $|r|=\beta_{i}$,
while at the same point either $r=s$ or $\eta(r)=0$, the quotients
are undefined. Fortunately, none of these conditions were observed.

\subsubsection{Extensions for the rectangular case}

\label{subsec:ExtRect}

We \textit{cannot} universally use the same approach here as for the
rhomboidal case (i.e. substituting $\mathrm{sgn}(w)$ for $w/|w|$),
because in the rectangular case $\partial_{s'}p$ consists of only
one term, according to Eqs. (\ref{eq:ds'p1}) and (\ref{eq:ds'p2}).
Thus, the quotient in Eq. (\ref{eq:dur}) is undefined if either $r=s$
or $\eta(r)=0$ and the quotient in Eq. (\ref{eq:dxr}) is undefined
if either $r=0$ or $|r|-\beta_{i}$. We can avoid this issue by working
at the level of the quotients, rather than the individual derivatives.

\paragraph{Extensions for shape function $p_{1}$\protect \protect \protect
\protect \\
 }

We note from Eqs. (\ref{eq:dsp1}) and (\ref{eq:ds'p1}) that $\partial_{s'}p_{1}=-\partial_{s}p_{1}$
and thus

\begin{equation}
q^{s}(p_{1})=\frac{\partial_{s}p_{1}}{\partial_{s'}p_{1}}=\frac{\partial_{s}p_{1}}{-\partial_{s}p_{1}}=-1\quad r\neq s,\eta(r)\neq0.\label{eq:qs1}
\end{equation}
The reasonable extension here is to ignore the restrictions, i.e.
always set 
\begin{equation}
q^{s}(p_{1})=-1\label{eq:qs1a}
\end{equation}

Next, we note from Eqs. (\ref{eq:ds'p1}) and (\ref{eq:dzp1}) that

\begin{equation}
\partial_{s'}p_{1}=\frac{1}{v_{h}}\frac{s-r}{|s-r'|}\partial_{z}p_{1}
\end{equation}
and thus

\begin{equation}
q^{z}(p_{1})=\frac{\partial_{z}p_{1}}{\partial_{s'}p_{1}}=\frac{\partial_{z}p_{1}}{\frac{1}{v_{h}}\frac{s-r}{|s-r'|}\partial_{z}p_{1}}=v_{h}\left(\frac{s-r}{|s-r|}\right)^{-1}=v_{h}\frac{s-r}{|s-r|},\;r\neq s.\label{eq:qz1}
\end{equation}
Then, by reasoning analogous to Section \ref{subsec:ExtRhom}, we
have

\begin{equation}
q^{z}(p_{1})=v_{h}\mathrm{sgn}(s-r).\label{eq:qz1a}
\end{equation}

\paragraph{Extensions for shape function $p_{2}$}

We note from Eqs. (\ref{eq:ds'p2}) and (\ref{eq:dsp2}) that

\begin{equation}
q^{s}(p_{2})=\frac{\partial_{s}p_{2}}{\partial_{s'}p_{2}}=\frac{0}{\partial_{s'}p_{i,2}}=0,\quad r\neq0,|r|\neq\beta_{i}.\label{eq:qs2}
\end{equation}
The reasonable extension here is to ignore the restrictions, i.e.
always set

\begin{equation}
q^{s}(p_{2})=0.\label{eq:qs2a}
\end{equation}

Next, we note from Eqs. (\ref{eq:ds'p2}) and (\ref{eq:dzp2}) that

\begin{equation}
q^{z}(p_{2})=\frac{\partial_{z}p_{i,2}}{\partial_{s'}p_{i,2}}=\frac{0}{\partial_{s'}p_{i,2}}=0,\quad r\neq0,|r|\neq\beta_{i}.\label{eq:qz2}
\end{equation}
Once again, the reasonable extension is to ignore the restrictions,
i.e. always set

\begin{equation}
q^{z}(p_{2})=0.\label{eq:qz2a}
\end{equation}

\subsection{Algorithmic Differentiation}

\label{subsec:AD}

Depending on the complexity of the expressions for the shift vector
components, deriving expressions for the corresponding partial derivatives
can be extremely time consuming, and may lead to unwieldy expreessions
that are difficult to check. This is the case for the partial derivatives
of $N_{y}$ in Lentz's original equations. However, these partial
derivatives are required to obtain the image shown in Fig. 4. (Recall
that partial derivatives of the shift vector are required to calculate
the Eulerian energy density.) Therefore, rather than deriving analytic
expressions using symbolic differentiation, \textit{Algorithmic Differentiation}
(AD) was used to obtain the required partial derivatives.\textcolor{blue}{{} }

\subsubsection{General principles of algorithmic differentiation}

AD differs from symbolic differentiation in several ways. 
\begin{itemize}
\item Symbolic differentiation: You manipulate formulas to obtain a new
symbolic expression for $f'(x)$. Once you have that expression, you
can plug in any $x$. For example, from $f(x)=\sin(x)+x^{2}$ you
derive $f'(x)=\cos(x)+2x$, then you evaluate at $x=2$. 
\item \textcolor{blue}{{} }Algorithmic Differentiation (AD): You never
build a closed-form expression for $f'(x)$. Instead: 
\begin{enumerate}
\item You record (or trace) the sequence of elementary operations that compute
$f(x)$. 
\item You propagate derivatives numerically through that sequence using
the chain rule. 
\item The chosen point $x$ is supplied before or at the start of the derivative
sweep; therefore, the chain rule is applied to actual numeric values,
not symbolic placeholders. Thus, AD produces a numeric derivative
at that point, not a symbolic formula for $f'(x)$. 
\end{enumerate}
\end{itemize}
\textcolor{blue}{{} }Standard symbolic differentiation can lead to
a much higher computational complexity than algorithmic differentiation.
This is illustrated in the following example. Consider a recursively
nested function 
\[
f_{n}(x)=\sin(\sin(\sin(\cdots\sin(x)\cdots))),
\]
where $\sin$ is applied $n$ times.\textcolor{blue}{{} } 
\begin{itemize}
\item Symbolic differentiation: By the chain rule, 
\[
f'_{n}(x)=\cos(f_{n-1}(x))\cdot f'_{n-1}(x).
\]
If expanded fully: 
\[
f'_{n}(x)=\cos(\sin(\sin(\cdots\sin(x))))\cdot\cos(\sin(\sin(\cdots\sin(x))))\cdot\cdots\cdot\cos(x),
\]
with $n$ nested cosine factors. For $n=3$: 
\begin{align*}
f_{3}(x) & =\sin(\sin(\sin(x))),\\
f'_{3}(x) & =\cos(\sin(\sin(x)))\cdot\cos(\sin(x))\cdot\cos(x).
\end{align*}
For large $n$, the expression becomes enormous: it involves $O(n^{2})$
trigonometric operations. \textcolor{blue}{{} } 
\item Algorithmic differentiation: At a chosen point $x$, AD computes numerically,
the following: 
\begin{align*}
v_{0} & =x,\quad d_{0}=1,\\
v_{k} & =\sin(v_{k-1}),\\
d_{k} & =\cos(v_{k-1})\cdot d_{k-1},\quad k=1,\dots,n.
\end{align*}
At the end, $f_{n}(x)=v_{n}$ and $f'_{n}(x)=d_{n}$. This requires
$O(n)$ operations. 
\end{itemize}
It should be noted that the above symbolic differentiation can be
accomplished more efficiently by re-using sub-expressions. In more
general cases, this approach is known as structured symbolic differentiation.
It can be a somewhat difficult and error-prone task to setup and code
an ad hoc structured symbolic differentiation algorithm. AD is just
a restricted, disciplined form of structured symbolic differentiation,
for which prepackaged tools have been created.\textcolor{blue}{{} }

Importantly, AD is \underline{not} equivalent to numerical methods
such as finite-difference algorithms\textcolor{blue}{.} If the function
to be differentiated is smooth at a given point, the result is highly
accurate. It does not suffer from round-off error and truncation errors,
unlike the finite differences method. In the latter, there is a trade-off
between these two types of error, such that the best one can hope
for is nearly single-precision accuracy in double-precision calculations.
However, AD is normally accurate to the full precision of the machine.

For the modified version of Lentz's warp drive, we ran a comparison
of the partial derivatives of the shift vector, as generated by AD
and by the analytic expressions, at all grid points, and found that
the results are in complete agreement.\footnote{This holds, except where extensions have not been applied in the AD
case (e.g. the planes $x=0$ and $y=0$). } This has given us confidence in applying AD to calculate these partial
derivatives for Lentz's original warp drive.

\textcolor{blue}{{} }

\subsubsection{AD for $N_{y}$ in Lentz's original warp drive}

Of course, one cannot expect AD to produce a meaningful result\footnote{Packaged AD applications may employ extensions similar to ours, but
that remains to be investigated.} for a particular partial derivative, if that derivative or the function
itself is undefined at the given point. As we have seen for the modified
Lentz geometry discussed in Section \ref{subsec:Ext}, the shift vector
has such points, which should be avoided when using AD.

As a practical matter, calculations of the shift vector components,
their derivatives and the Eulerian energy density are performed at
a finite number of locations, which are at points on a regularly-spaced
grid. The grid is set up such that the $x$-, $y$-, and $z$-axes
coincide with grid lines, and the spacing of the grid is such that
all centerlines of the rhomboids coincide with grid lines. As discussed
in Section \ref{subsec:phiLentz}, we avoided possible ambiguities
at $y=0$ by taking the slice\textcolor{blue}{s} at $y=10^{-6}$.
This was implemented by shifting the entire grid by $10^{-6}$, along
the $y$ direction.

\section{Estimating Total Energy}

\label{sec:EstEtot}

As discussed in section \ref{subsec: Energy}, the energy density
at a point in space is the sum of what we call the $U$ and the $T$
terms.

Based on the assumption that the energy density at a grid point is
approximately constant over the grid cell containing that point, the
total contribution from the $U$ term is estimated by adding the energy
density at all grid points and multiplying by the cell volume, i.e.

\begin{equation}
U_{tot}\approx\Delta^{3}\sum_{i}\sum_{j}\sum_{k}E_{i,j,k}
\end{equation}
where $\Delta$ denotes the fixed grid spacing.

The total contribution from the $T$ term is estimated by adding up
the contributions at all grid points in the planes $x=0$ and $y=0$.
(The $T$ term is zero at all other grid points). We now derive the
approximate expression for the contribution of the $T$ term to the
total energy. Based on the definition of $T$, the integral over all
space of $T$ is given by

\begin{eqnarray}
T_{tot} & = & 2\int_{z=-\infty}^{\infty}\int_{y=-\infty}^{\infty}\int_{x=-\infty}^{\infty}\delta(x)A(x,y,z)dxdydz\nonumber \\
 & + & 2\int_{z=-\infty}^{\infty}\int_{y=-\infty}^{\infty}\int_{x=-\infty}^{\infty}\delta(y)A(x,y,z)dxdydz\nonumber \\
 & + & 4\int_{z=-\infty}^{\infty}\int_{y=-\infty}^{\infty}\int_{x=-\infty}^{\infty}\delta(x)\delta(y)B(x,y,z)dxdydz\label{eq:IofT}
\end{eqnarray}
where

\begin{equation}
A=\partial_{s}\phi\left(\partial_{s}^{2}\phi+\partial_{z}N_{z}\right)
\end{equation}
and

\begin{equation}
B=\left(\partial_{s}\phi\right)^{2}.
\end{equation}
From the distributional properties of the deltas, we can immediately
rewrite Eq. (\ref{eq:IofT}) as

\begin{equation}
T_{tot}=2\int_{z=-\infty}^{\infty}\int_{y=-\infty}^{\infty}A(0,y,z)dydz+2\int_{z=-\infty}^{\infty}\int_{x=-\infty}^{\infty}A(x,0,z)dxdz+4\int_{z=-\infty}^{\infty}B(0,0,z)dz
\end{equation}
and using the same reasoning regarding the grid as before, we can
estimate $T_{tot}$ via

\begin{equation}
T_{tot}\approx2\Delta^{2}\sum_{i}\sum_{j}A(0,y_{j},z_{i})+2\Delta^{2}\sum_{i}\sum_{k}A(x_{k},0,z_{i})+4\Delta\sum_{i}B(0,0,z_{i}).
\end{equation}

It is expected that the estimates of $U_{tot}$ and $T_{tot}$ improve
as $\Delta$ shrinks.

\section{ A counterexample to the shift-derivative inequality}

\label{sec:New-SDineq}

Our counter-example will use the function 
\begin{equation}
\rho(\alpha,s')=(1-|s'|)e^{-\alpha^{2}}.
\end{equation}
Set $\alpha=(z-|s-s'|)$. Then $\partial_{z}\rho(\alpha,s')=-2(z-|s-s'|)\rho(\alpha,s')$.
Set $z=0$ and $s>0$. Then Eqs. (\ref{Eq:|dzNzx|}) become, 
\begin{equation}
\begin{aligned}|\partial_{z}^{2}\phi| & =\frac{1}{2}\left|\int_{-\infty}^{\infty}ds'(1-|s'|)|s-s'|e^{-(s-s')^{2}}\right|\\
|\partial_{z}\partial_{s}\phi| & =\frac{1}{2}\left|\int_{-\infty}^{\infty}ds'\text{sgn}(s-s')(1-|s'|)|s-s'|e^{-(s-s')^{2}}\right|
\end{aligned}
\label{Eq:|dzNzs|1}
\end{equation}
These integrals can be computed analytically by first dividing them
into three regions. Define $I_{1}$, $I_{2}$ and $I_{3}$ as 
\begin{equation}
\begin{aligned}I_{1} & =\int_{s}^{\infty}ds'(1-s')(s'-s)e^{-(s-s')^{2}}\\
I_{2} & =\int_{0}^{s}ds'(1-s')(s-s')e^{-(s'-s)^{2}}\\
I_{3} & =\int_{-\infty}^{0}ds'(1+s')(s-s')e^{-(s'-s)^{2}}
\end{aligned}
\end{equation}
We then compute the integrals and from Eq. (\ref{Eq:|dzNzs|2}) we
have 
\begin{equation}
\begin{aligned}|\partial_{z}^{2}\phi| & =\frac{1}{2}\left|I_{1}+I_{2}+I_{3}\right|=\frac{1}{2}\left|1-s-\frac{1}{2}\Gamma\left(\frac{1}{2},s^{2}\right)\right|,\\
|\partial_{z}\partial_{s}\phi| & =\frac{1}{2}\left|-I_{1}+I_{2}+I_{3}\right|=\frac{1}{2}\left|\frac{\sqrt{\pi}}{2}-\frac{1}{2}\Gamma\left(\frac{1}{2},s^{2}\right)\right|,
\end{aligned}
\label{Eq:|dzNzs|2}
\end{equation}
where the $\Gamma$'s are incomplete gamma functions. For example,
assume $s=1$. Then $|\partial_{z}^{2}\phi|(1,0,0)=0.07$ and $|\partial_{z}\partial_{s}\phi|(1,0,0)=0.4$.
Therefore $|\partial_{z}^{2}\phi|(1,0,0)<|\partial_{z}\partial_{s}\phi|(1,0,0)$,
thereby contradicting the shift-derivative inequality.\footnote{If we change $s$ and $s'$ to $x$ and $x'$ then the shift-derivative
inequality becomes equivalent to Lentz's inequality in Ref. \cite{Lentz}.}

\section{Graphics}

\label{sec:E-graphics}

Two types of graphical representations of the quantities of interest
are used in the figures shown in this paper: 
\begin{itemize}
\item \textit{A Planar cross-section} provides an excellent depiction of
the spatial structure of a quantity, but only a vague sense of the
numerical values of this quantity, as it varies over the plane. This
is because the numerical values are conveyed using color intensity,
for which the human eye is not well-adapted to discern differences.
(More about this below.) 
\item \textit{A Profile} provides an excellent depiction of the numerical
values of a quantity, as it varies along a specific line, but being
confined to that specific line, it does not convey much information
about spatial structure. 
\end{itemize}
The two types of figure are discussed in Appendices \ref{subsec:E.1}
and \ref{subsec:E.2}. In both cases, graphical representations are
created from quantities that computed at all points on a dense, 3-dimensional
grid. However, the grid is constructed to be compatible with the image
format used for planar cross-sections. The grid structure is also
discussed in Appendix \ref{subsec:E.1}.

\subsection{Planar cross-sections}

\label{subsec:E.1}

The 3-dimensional grid on which quanties are computed has the following
properties: 
\begin{itemize}
\item It covers a finite volume of a space-like hypersurface, which is represented
Cartesian coordinate system. 
\item This volume has the shape of a cube centered a the origin of the Cartesian
coordinate system. 
\item The cube is oriented such that each Cartesian axis intersects the
center point of a pair of opposite faces. 
\item Grid points are equally spaced along the Cartesian axes, and along
all lines parallel to them, each of which passes through one of the
former. 
\end{itemize}
We define a planar cross-section as a 2-dimensional slice of the grid,
which is orthogonal to one of the Cartesian axes (and thus parallel
to the plane defined by the other two axes) and centered at one of
the grid points on the orthogonal axis.

However, the appropriate resolution of the grid remains to be determined.
Planar cross-sections are displayed in a square image, consisting
of $301\times301$ pixels. An odd number of pixels is chosen so that
the center of the cross-section can be exactly centered in the image
(with 150 pixels on either side of that origin in each dimension).
The total number of pixels (301) across the image was chosen to be
of sufficient density to create the illusion of continuity across
the image, despite the fact that the data are generated at discrete
locations. Thus, to achieve maximum resolution, the 3-dimensional
grid must be $301\times301\times301$ points, so that a planar cross-section
of the grid may be put into 1-to-1 correspondence with the pixels
in the image. At this resolution there are $301^{2}=90,601$ points
on each planar cross-section and $301^{3}=27,270,901$ points on the
entire grid.

For each pixel in the image, we must supply a color and intensity
that, together, represents a numerical value of the quantity. This
graphical representation is implemented using OpenGL, which requires
that the real values of the color intensity be in the range $[0,1]$,
where the intensity scale is linear. That is, an intensity of 1 corresponds
to maximum brightness and an intensity of 0.5 is exactly half as bright,
0.25 a quarter as bright, etc., with 0 intensity being black. The
linearity of intensity exacerbates the problem mentioned above (that
the human eye is not well-adapted to discern differences in intensity).
This is because the human eye does not perceive brightness linearly.
In particular, dark regions with intensities a high as 0.15 are perceived
as black, that is, indistinguishable from zero intensity.

\subsubsection{$\gamma$-correction}

\label{subsec:E.1.1}

A well-established remedy for the problem described in the preceding
paragraph is $\gamma$-correction, in which intensity values are modified
according to the formula $V_{\mathrm{out}}=V_{\mathrm{in}}^{\gamma}$,
where it is required that $V_{\mathrm{in}}\geq0$ and $\gamma\in[0,1]$.
Thus, $\gamma=1$ implies no $\gamma$-correction and the smaller
the $\gamma$ the greater the correction. Note that, for $V_{\mathrm{in}}\in[0,1]$
as required by OpenGL, if some non-trivial actual $\gamma$-correction
is applied, that is, $\gamma<1$, it will have no effect at the top
of the scale, that is, $1^{\gamma}=1$, and has increasing impact
as $V_{\mathrm{in}}\rightarrow0$, shifting values towards higher
intensity, such that the smaller the value $V_{\mathrm{in}}$, the
greater the shift. For example, suppose $\gamma=0.5$: if $V_{\mathrm{in}}=0.5$,
then $V_{\mathrm{in}}^{\gamma}=0.7=1.4V_{\mathrm{in}}$, whereas if
$V_{\mathrm{in}}=0.25$, then $V_{\mathrm{in}}^{\gamma}=0.5=2V_{\mathrm{in}}$,
and if $V_{\mathrm{in}}=0.125$, then $V_{\mathrm{in}}^{\gamma}=0.35=2.8V_{\mathrm{in}}$.

Note that too much $\gamma$-correction, that is, very small values
of $\gamma$, is counter-productive, as it tends to make all points
equally bright. Thus, although no details are hidden, there appears
to be little or no variability in the intensity.

\subsubsection{Algorithm for the conversion of computed values to color intensities}

\label{subsec:E.1.2}

Based on the preceding discussion, a computed value $V_{c}$ of a
physical quantity is processed as follows: 
\begin{enumerate}
\item $V_{c}$ is normalized by the largest absolute value, among all computed
values of that quantity, i.e. $V_{n}=V_{c}/|V_{c}|_{\max}$. 
\item The desired $\gamma$-correction is applied, that is, $V_{n,\gamma}=\left|V_{n}\right|^{\gamma}$. 
\item If $V_{n}<0$, then $V_{n,\gamma}$ is passed to OpenGL, as the red
intensity; otherwise, it is passed to OpenGL as the green intensity. 
\end{enumerate}
Owing to the normalization step, the final graphical representation
of a planar cross-section cannot convey any information about the
true range of values. Therefore, each planar cross-section is accompanied
by a legend (on the right) that shows the true range of values.

\subsection{Profiles}

\label{subsec:E.2}

A profile can be obtained for any line on the planar cross-section
for a particular quantity. Thus, lines parallel to either edge of
the cross-section, that is, parallel one Cartesian axis, will have,
at most, 301 points. (These are the only types of profiles used in
this study.) This is sufficient to provide a continuous appearance
to the resultant graphs, at the fixed, chosen size for those graphs.
Thus, the graph shows the actual numerical values (not the normalized
values) of a particular, computed quantity, as it varies along the
chosen line, with the computed values as the ordinates and values
of one Cartesian coordinate as the abscissae. The graphs were produced
using GNUplot.

\section{Superluminal travel}

\label{sec:F-superlumin}

Alcubierre \cite{Alcubierre} introduced warp drives as geometries
which permit an observer to move in free fall at superluminal speeds
relative to asymptotic time, and whose proper time is not dilated
relative to the asymptotic time. In this appendix, we demonstrate
that these properties can be obtained from the geometry generated
by a hyperbolic potential (HP). In particular, we show that any HP
with velocity $\textbf{v}$ can be scaled so that it permits an observer
to move in free fall at velocity $\textbf{v}$ and without time dilation.
The analysis closely follows that of Shoshany \cite{Shoshany}.

Recall the metric given by Eq. (\ref{Eq:metric}) 
\begin{equation}
\begin{aligned}ds^{2} & =-(1-N^{i}N_{i})dt^{2}-2N_{i}dx^{i}dt+\sum_{i=1}^{3}dx^{i2}\\
 & =-dt^{2}+\sum_{i}\left(N_{i}dt-dx^{i}\right)^{2}.
\end{aligned}
\end{equation}
Note that if $\tau$ is the proper time, and if an object follows
the trajectory $\left(t(\tau),\textbf{x}(\tau)\right)$ such that
$\textbf{x}(\tau)=\textbf{x}_{0}+\textbf{N}t(\tau)$, then $\left(N_{i}dt-dx^{i}\right)=0$
so that $ds^{2}=-dt^{2}$. However, for a timelike path, the proper
time is just $\sqrt{-ds^{2}}$, so $dt=d\tau$. Thus, such an object
does not exhibit time dilation.

Is such an object in free fall? In other words, is the object's trajectory
a geodesic? A geodesic is an extremal of the proper time 
\begin{equation}
\begin{aligned}\tau & =\int\mathcal{L}d\lambda\\
 & =\int\sqrt{-g_{\mu\nu}\dot{x}^{\mu}\dot{x}^{\nu}}d\lambda,
\end{aligned}
\end{equation}
where $\dot{x}^{\alpha}$ denotes $\frac{dx^{\alpha}}{d\lambda}$.
The integral can be shown to be independent of the parameter $\lambda$
so we can take $\lambda=\tau$. For our geometry, 
\begin{equation}
ds^{2}=g_{\mu\nu}dx^{\mu}dx^{\nu},
\end{equation}
and the values of $g_{\mu\nu}$ are given by 
\begin{equation}
\begin{aligned}g_{00} & =-1+N_{i}N^{i}\\
g_{0i} & =-N_{i}\\
g_{ij} & =\delta_{ij}.
\end{aligned}
\end{equation}

To find extrema, we solve the Euler-Lagrange equation 
\begin{equation}
\frac{d}{d\tau}\left(\frac{\partial\mathcal{L}}{\partial\dot{x}^{\alpha}}\right)-\frac{\partial\mathcal{L}}{\partial x^{\alpha}}=0.
\end{equation}

We have 
\begin{equation}
\begin{aligned}\frac{\partial{\mathcal{L}}}{\partial\dot{x}^{\alpha}}=\frac{-g_{\alpha\mu}\dot{x}^{\mu}}{\sqrt{-g_{\mu\nu}\dot{x}^{\mu}\dot{x}^{\nu}}}\end{aligned}
\end{equation}
and 
\begin{equation}
\begin{aligned}\frac{\partial{\mathcal{L}}}{\partial x^{\alpha}}=\frac{-\partial_{\alpha}g_{\mu\nu}\dot{x}^{\mu}\dot{x}^{\nu}}{2\sqrt{-g_{\mu\nu}\dot{x}^{\mu}\dot{x}^{\nu}}}\end{aligned}
\end{equation}

Consider an ansatz for the solution of these E-L equations. 
\begin{equation}
\left(\dot{t},\dot{\textbf{x}}\right)=(1,\textbf{N}).
\end{equation}
Note that if observers have a 4-velocity $(1,\textbf{N})$, they are
Eulerian observers. Start with the time component. 
\begin{equation}
\begin{aligned}\frac{d}{d\tau}\frac{\left((1-N^{2})+N^{2}\right)}{\sqrt{\left((1-N^{2})+2N^{2}-N^{2}\right)}}-\frac{-2\partial_{\alpha}N_{i}N^{i}+2\partial_{\alpha}N_{i}N^{i}}{2\sqrt{\left((1-N^{2})+2N^{2}-N^{2}\right)}} & =\frac{d}{d\tau}(1)-0=0.\end{aligned}
\end{equation}

The spacial-components are similar. 
\begin{equation}
\begin{aligned}\frac{d}{d\tau}\frac{\left(N_{i}-N_{i}\right)}{2\sqrt{\left((1-N^{2})+2N^{2}-N^{2}\right)}}-\frac{-\left(2N^{j}\partial_{i}N_{j}-2N^{j}\partial_{i}N_{j}\right)}{2\sqrt{\left((1-N^{2})+2N^{2}-N^{2}\right)}}=0.\end{aligned}
\end{equation}
Therefore, we established that our ansatz is a solution to the E-L
equations. From this it follows that 
\begin{equation}
\frac{d\textbf{x}}{dt}=\textbf{N}.
\end{equation}

In summary, we have shown that if an object travels with a velocity
$\textbf{N}$, then it is in free-fall, and its proper time is $t$;
that is, it experiences no time dilation. Interestingly, we have nowhere
referred to the velocity of the warp drive, which we'll call $\textbf{v}'$.
If the free-falling object's velocity $\textbf{N}$ is different from
$\textbf{v}'$, then even if it starts off in the center of the warp
drive (however we define that), it will move away from the center
and eventually towards regions of the warp drive with ever smaller
values of $\textbf{N}$. If we want to arrange a situation where the
freely-falling object remains in the center of the warp drive --
and thus continues to experience the same value of $\textbf{N}$ and
therefore the absence of time dilation -- then we must have $\textbf{v}'=\textbf{N}$.
For the HP's, we can control the value of $N$ by modifying various
parameters, so we can set up a warp drive with a velocity $\textbf{v}'$
and, at its central point, $\textbf{v}'=\textbf{N}$. An object moving
freely in the center of the warp drive will continue to stay in the
center of the drive, and will experience no time dilation. If the
region around the center of the warp drive has a level shift vector
(i.e. $\textbf{N}$ is constant in that region), then an object moving
(by non-gravitational forces) within that central region will continue
to stay in that region.

\section{The y-dependence of the original and the modified Lentz warp drives}

\label{sec:G-yDep}

Lentz's original warp drive is based on the potential $\phi_{L}$,
which we observed to be asymmetric in the exchange of $x$ and $y$.
The source function and shift vectors for $y=0$ should be the same
(up to a certain sign) for $\phi_{L}$ (Eq. (\ref{Eq:de})) geometry
and $\phi_{\text{rh}}$ (Eq. (\ref{Eq:phiMod})). However, the two
geometries differ, away from $y=0$. The Eulerian energy density at
$y=0$ is dependent on the $y$-component of the shift vector as well
as the $y$-dependence of all components of the shift vector. Therefore
the Eulerian energy density is different for the $\phi_{L}$ geometry
than for the $\phi_{\text{rh}}$ geometry.

Here we show $N_{y}^{\text{rh}}$ and $N_{y}^{\text{L}}$\textcolor{green}{.}
Notice that these exhibit discontinuities around the $y$-axis.

\begin{figure}[htbp]
\centering \begin{subfigure}[b]{0.45\textwidth} \centering \includegraphics[width=1\textwidth]{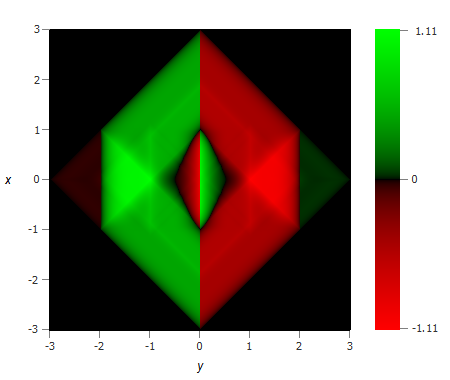}
\caption{$N_{y}^{L}$}
\label{Fig:Ny_L_rhom} \end{subfigure} \hfill{}\begin{subfigure}[b]{0.45\textwidth}
\centering \includegraphics[width=1\textwidth]{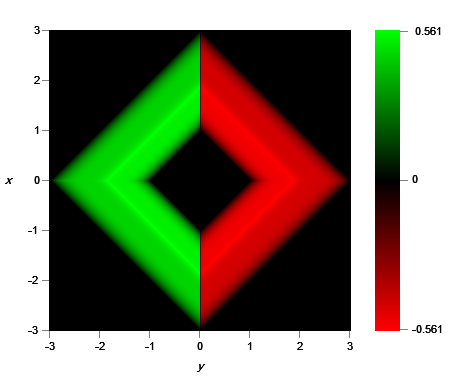} \caption{$N_{y}^{\text{rh}}$}
\label{Fig:Ny_rh_rhom} \end{subfigure} \caption{The shift vector components evaluated on the slice $z=0$. The graphics
gamma correction value (see Appendix\textcolor{green}{{} }%
\mbox{%
\ref{subsec:E.1.1}%
}) is set to\textcolor{green}{{} }%
\mbox{%
$\gamma=0.5$%
}.}
\label{Fig:Ny_rh} 
\end{figure}

\begin{figure}[H]
\centering \includegraphics[scale=0.42]{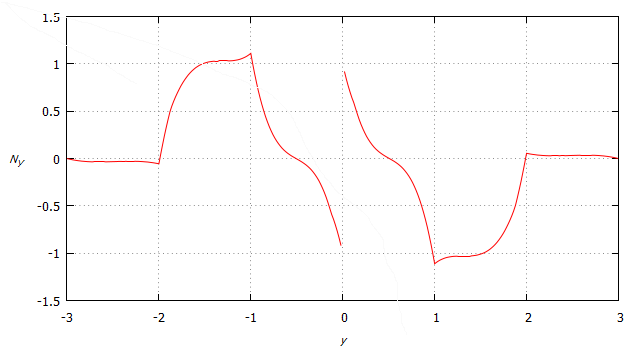} \caption{The profile $N_{y}^{\text{L}}$ at fixed $(z,x)=(0,0).$}
\label{Fig:Ny_L_rhom_pfl} 
\end{figure}

We show further detail on the $N_{y}^{\text{rh}}$ discontinuity\textcolor{green}{,}
in Fig. \ref{Fig:Ny_L_rhom_pfl} where we plot the profile $N_{y}^{\text{rh}}(x,y,z)$
at a single ($x,z$) value. The figure also shows the antisymmetry
of $N_{y}^{\text{rh}}$ around $y=0$. 

\section{Boundedness of the Shift Vector Domain}

\label{sec:H-Boundedness}

For our proof of finiteness, we repeat the conditions described in
Section \ref{subsec:Finiteness}: 
\begin{itemize}
\item The linear-cancellation condition

\begin{equation}
\int_{-\infty}^{\infty}ds'\rho(s',\alpha\pm s')=0\label{Eq:H-1}
\end{equation}
for any constant $\alpha$. 
\item The source-domain boundedness condition 
\begin{equation}
\begin{aligned} & \rho\;\text{has a bounded domain such that}\;\rho(s,z)=0\\
 & \;\text{if}\;|z|>B\;\text{or}\;|s|>B
\end{aligned}
\label{Eq:H-2}
\end{equation}
for some $B$. 
\end{itemize}
Our proof\textcolor{magenta}{{} }in Appendix \ref{subsec:H.1} will
also depend on the definition $s(x,y)=|x|+|y|$, from which we can
conclude that $s>0$ and from which we can also conclude (which we
use at the end of our proof) that if $|x|>B$ or $|y|>B$, then $s>B$.
We also show in Appendix \ref{subsec:H.2} that if $s(x,y)=x+y$,
no such proof is possible. Thus, for these definitions of $s$, the
domain of the shift vector components is not bounded; therefore, the
energy is not finite.

\subsection{Boundedness of the shift vector domain when $s(x,y)=|x|+|y|$}

\label{subsec:H.1}

From our expression Eq. (\ref{Eq:phiMod}) for the potential and the
definition of the shift-vector, we obtain the shift components (again
setting $v_{h}=1$) 
\begin{equation}
\begin{aligned}N_{x} & =\partial_{x}\phi=\frac{1}{4}\mathrm{sgn(x)}\int_{-B}^{B}ds'\mathrm{sgn}(s-s')\rho(s',z-|s-s'|),\\
N_{y} & =\partial_{y}\phi=\frac{1}{4}\mathrm{sgn}(y)\int_{-B}^{B}ds'\mathrm{sgn}(s-s')\rho(s',z-|s-s'|),\\
N_{z} & =\partial_{z}\phi=-\frac{1}{4}\int_{-B}^{B}ds'\rho(s',z-|s-s'|),
\end{aligned}
\label{Eq:H-3}
\end{equation}
where we use the bounds $\pm B$ as the limits of integration (rather
than $\pm\infty)$.

To show the boundedness of the $N_{i}$ domain, it suffices to find
$Z_{b},X_{b},Y_{b}$ so that $N_{i}(x,y,z)=0$ unless $|z|\le Z_{b}$,
and $|x|\le X_{b}$ and $|y|\le Y_{b}$. 
\begin{itemize}
\item Large $z$. 
\begin{itemize}
\item First, consider $z\ge4B$ and $s>2B$. Examine the integrand(s). We
know that $s'\le B$ so $|s-s'|=s-s'$ and $z-|s-s'|=z-s+s'$. Also
$\mathrm{sgn}(s-s')=1$. The $N_{i}$ are then proportional to $\int_{-\infty}^{\infty}ds'\rho\left(s',(z-s)+s'\right)$.
We see this is of the form of the linear-cancellation condition and
therefore $N_{i}=0$. 
\item Next, consider $z\ge4B$ and $0\le s\le2B.$ Since $-B\le s'\le B$,
we can derive $-B\le(s-s')\le3B$ and therefore $|s-s'|\le3B$. From
this, we get $z-|s-s'|\ge4B-3B=B$. Therefore, in the integrands we
have $\rho(s',z-|s-s'|)=0$ since the second argument is greater than
$B$. Hence, again, $N_{i}=0$. 
\end{itemize}
\item Large $-z$. Consider $z\le-3B$. But then $z-|s-s'|<-3B$, so the
second argument of $\rho$ (in the integrand) is less than $-B$ so
$\rho(s',z-|s-s'|)=0$ for all $s'$ and therefore $N_{i}=0$. 
\item Large $s$. Consider $s\ge B$. Then, since $s'\le B$, we have $|s-s'|=s-s'$
and $\mathrm{sgn}(s-s')=1$ almost everywhere.\footref {note1}
Then $z-|s-s'|=z-s+s'$ and we can, as before, invoke the linear-cancellation
condition to see that $N_{i}=0$. 
\end{itemize}
If we select $Z_{b}=3B$, $X_{b}=B$ and $Y_{b}=B$, then if $|z|>Z_{b}$,
or $|x|>X_{b}$ or $|y|>Y_{b}$, then $|z|>Z_{b}$ or $|s|>B$ and
thus we have proven the boundedness of the $N_{i}$ domains.

\subsection{Why the energy could be infinite if $s=x+y$}

\label{subsec:H.2}

Previously, we pointed out that $\phi$ is a solution to Eq. (\ref{Eq:de})
\textit{almost everywhere}\footref{note1} -- but not everywhere.
Instead, $\phi$ is a solution of Eq. (\ref{Eq:DE}), which differs
from Eq. (\ref{Eq:de}) by a term $S(x,y,z)$ originating from derivatives
of the absolute values appearing in the definition of $s(x,y)$.

This naturally leads us to consider possible solutions $\phi$ that
depend on $s(x,y)=x+y$. Because that definition of $s$ doesn't involve
absolute values, then the term $S$ is absent.

Unfortunately, that definition of $s$ does not lead to boundedness
of the shift-vectors and, therefore, finiteness of the energy. We
can easily see this by examining our proof of boundedness for the
definition $s=|x|+|y|$. The proof demonstrates that when either $|z|$
or $s$ is large, the shift vector is 0. However, boundedness is with
respect to the coordinates $x,y,z$. When $s=|x|+|y|$, then $s$
is large if either $|x|$ or $|y|$ are large. However, when $s=x+y$,
we can have large $x$ with $y\approx-x$ and therefore $s\approx0$.
Our boundedness-proof has nothing to say about situations where both
$s$ and $z$ are small; therefore, in general, we can't show, for
the case $s=x+y$ that the shift-vectors are $0$ if either $|x|$
or $|y|$ are large.

For the rhomboidal-source $\rho_{\text{rh}}$, we now show that with
$s=x+y$, the shift-vector would indeed have an unbounded domain and
therefore, the energy would be infinite. Our source would now be 
\begin{equation}
\rho_{\text{rh}}^{(s=x+y)}(s',z')=\rho_{\text{rhomb}}(x'+y',z').
\end{equation}
If we substitute that source in Eq. (\ref{Eq:phiMod}), then we find
\begin{equation}
N_{x}^{(s=x+y)}=\partial_{x}\phi=\frac{1}{4}\int_{-B}^{B}ds'\mathrm{sgn}((x+y)-s')\rho_{\text{rh}}^{(s=x+y)}(s',z-|(x+y)-s'|).
\end{equation}
Let $|x|$ be arbitrarily large, but $x=-y+2$. Then 
\begin{equation}
N_{x}^{(s=x+y)}=\partial_{x}\phi=\frac{1}{4}\int_{-B}^{B}ds'\mathrm{sgn}(2-s')\rho_{\text{rh}}^{(s=x+y)}(s',z-|2-s'|).
\end{equation}
$N_{x}^{(s=x+y)}(s,z)$ is obtained from Fig. (\ref{Fig:Nx_rhom}),
where $x$ is substituted for $s$. When $x=-y+2$ and $z=0$ , the
coordinate $(s,z)=(2,0)$. We see in Fig. (\ref{Fig:Nx_rhom}) that
$N_{x}(2,0)$ is non-zero and thus, as claimed, the shift-vector component
has an unbounded domain.

\clearpage\bibliography{Warp}
 \bibliographystyle{IEEEtran} 
\end{document}